\documentclass[%
 reprint,
superscriptaddress,
nofootinbib,
]{revtex4-2}
\usepackage{amsmath}
\usepackage{amssymb}
\usepackage{amsfonts}
\usepackage{graphicx} 
\usepackage{amsfonts}
\usepackage{xcolor}
\usepackage{mathtools}
\usepackage{graphicx}
\usepackage{dcolumn}
\usepackage{braket}
\usepackage{amsthm}
\usepackage{booktabs}
\usepackage{subcaption}

\newcommand{\SupI}{\emph{Supplementary Information}}

\newcommand{\lc}{\mathcal{L}}

\newcommand{\alcomm}[1]{\textcolor{blue}{#1}}
\newcommand{\ajcomm}[1]{ \hfill \break  \{ \textcolor{blue}{AJ:#1} \}  }
\newcommand{\ctcomm}[1]{\textcolor{red}{CT:~#1}}

\long\def\ca#1\cb{} 
\newif\ifshowtext
\showtextfalse 

\newcommand{\conditionaltext}[1]{%
    \ifshowtext
        #1%
    \fi
}

\begin{document}


\title{Learning response functions of analog quantum computers:\\ analysis of neutral-atom and superconducting platforms}

\author{Cenk T\"uys\"uz}
\affiliation{Theoretical Division, Los Alamos National Laboratory, Los Alamos, New Mexico 87545, USA}
\affiliation{Deutsches Elektronen-Synchrotron DESY, Platanenallee 6, 15738 Zeuthen, Germany}
\affiliation{Institut für Physik, Humboldt-Universität zu Berlin, 12489 Berlin, Germany}
\author{Abhijth Jayakumar}
\email{abhijithj@lanl.gov}
 \affiliation{Theoretical Division, Los Alamos National Laboratory, Los Alamos, New Mexico 87545, USA}
\author{Carleton Coffrin}
 \affiliation{Advanced Network Science Initiative, Los Alamos National Laboratory, Los Alamos, New Mexico 87545, USA}
\author{Marc Vuffray}
 \affiliation{Theoretical Division, Los Alamos National Laboratory, Los Alamos, New Mexico 87545, USA}
\author{Andrey Y. Lokhov}
\email{lokhov@lanl.gov}
 \affiliation{Theoretical Division, Los Alamos National Laboratory, Los Alamos, New Mexico 87545, USA}

\begin{abstract}
Analog quantum computation is an attractive paradigm for the simulation of time-dependent quantum systems. Programmable analog quantum computers have been realized in hardware using a variety of physical principles, including neutral-atom and superconducting technologies. The input parameters of the physical Hamiltonians that are used to program the quantum simulator generally differ from the parameters that characterize the output distribution of data produced under a specified quantum dynamics. The relationship between the input and output parameters is known as the response function of the analog device. Here, we introduce a streaming algorithm for learning the response function of analog quantum computers from arbitrary user inputs, thus not requiring special calibration runs. We use the method to learn and compare the response functions of several generations of analog quantum simulators based on superconducting and neutral-atom programmable arrays.
\end{abstract}

\maketitle


\section{Introduction}

The paradigm of analog quantum computation relies on the idea of implementing the hardware analogy of a quantum system that one would like to study \cite{wurtz2023aquila, henriet_quantum_2020,dwave-advantage,flannigan2022propagation,scholl2021quantum,browaeys2020many}. Continuous physical parameters describing a quantum system, such as interactions between quantum particles, are then exposed as programmable variables available to a user. This is in contrast to digital quantum computation paradigm that relies on discrete qubits and gates and requires an extra layer of programming abstraction to translate physical problems into executable quantum circuits \cite{kwon2021gate,gonzalez2021scaling}.

By their nature, analog quantum computers straddle the line between experiment and computing, making these devices promising candidates for the simulation of large-scale quantum systems \cite{daley2022practical}. Moreover, depending on the mode of dynamic evolution, analog quantum computers also promise to advance the solution to a variety of problems beyond quantum simulation \cite{tasseff2022emerging,pang2021potential,ebadi2022quantum,dixit2021training}. For example, a popular class of dynamic evolution is given by quantum annealing, an algorithm inspired by the ideas from adiabatic quantum computation \cite{farhi2000quantum}.

Due to the peculiarities of the quantum dynamics, programming errors, quantum decoherence, and inevitable analog noise and biases, the output parameters characterizing the quantum state realized in analog computing devices at the end of the quantum evolution can be different from the values expected from the programmed input values \cite{pearson2019analog,perdomo2016determination,marshall2017thermalization}. The relation between the input and output parameters is often referred to as a \textit{response function} of a device. The knowledge of the response function enables the user to predict the computation outcome, can provide valuable information about the internal properties of an analog quantum device such as qubit noise \cite{vuffray2022programmable, nelson2021single, suau2023single, morrell2023signatures}, and can be used for the calibration of analog machines.

\begin{figure*}
    \centering
    \begin{minipage}[c]{.47\textwidth}
    \vfill
    \subfloat[]{\includegraphics[scale=0.39]{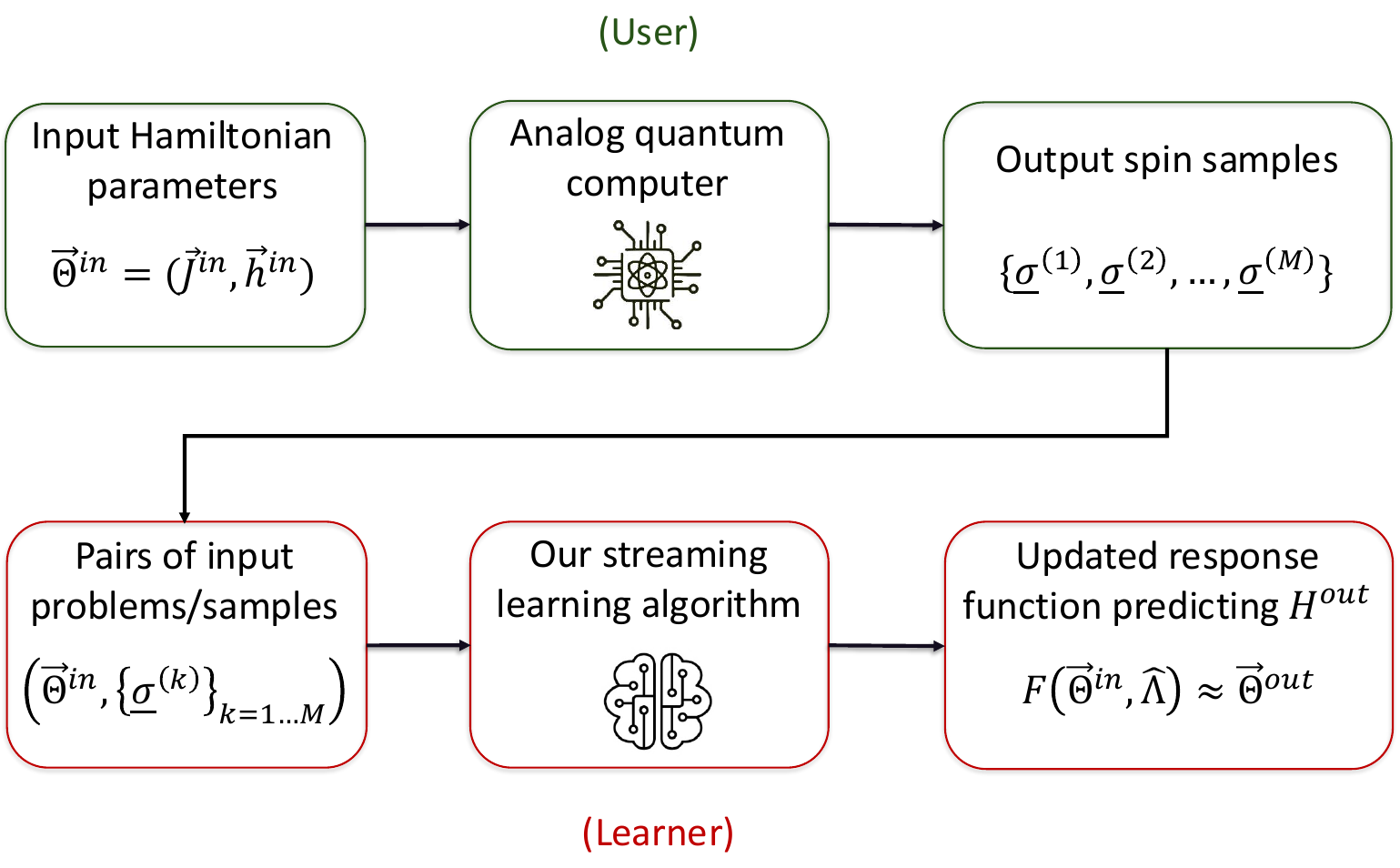}}
    \end{minipage}
    \vspace{0.1in}
    \hfill
    \begin{minipage}[c]{.47\textwidth}
    \vfill
    \subfloat[]{\includegraphics[scale=0.18]{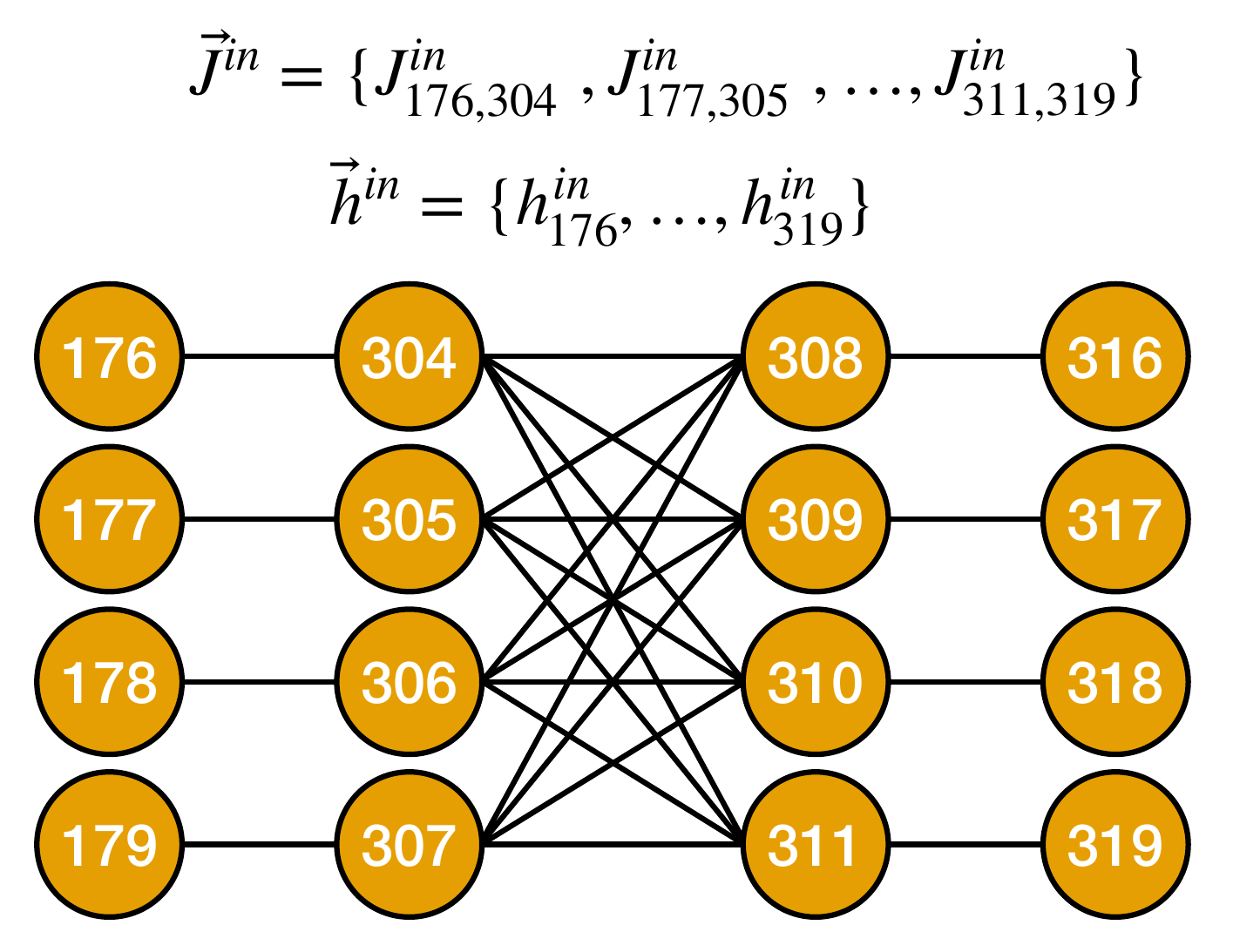}}\\\vfill
    \subfloat[]{\includegraphics[scale=0.3]{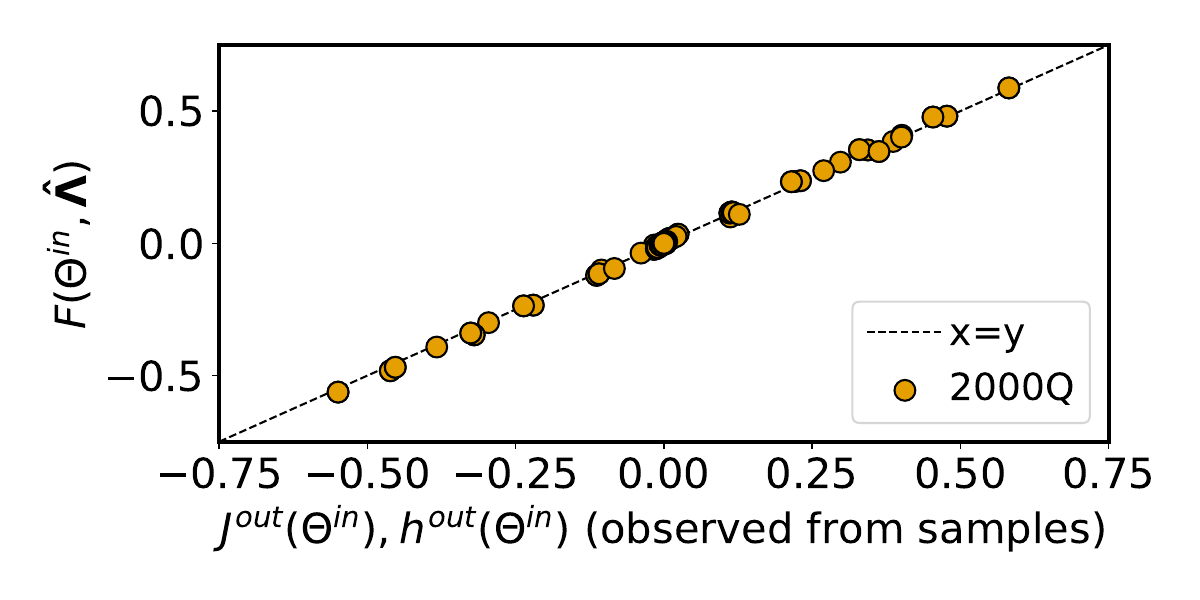}}\\\vfill
    \end{minipage}
\caption{{\bf Setup for the online learning of the response function.} (a) The users execute unrelated computational tasks on the analog quantum computers, producing sequence of different input Hamiltonian parameters and arbitrary sets of output samples collected at the end of the quantum evolution for each of the input problems. These pairs of input parameters and samples are then passed to the streaming algorithm that learns and refines the parameters of the response function in an online function. The algorithm can operate in the background during system use and refine potentially drifting response parameters in real-time.
(b) Illustration of our approach on learning a response function for the 16-qubit portion of the D-Wave 2000Q superconducting analog quantum computing platform (qubit numbers in the figure). The training set consists of 225 of random input Hamiltonian parameters in the range with couplings in the range [-0.05, 0.05] . Here, using a test set of 5 new input Hamiltonians and a sufficiently large number of samples collected for each of them, we compare the output parameters predicted by the learned response function ($F_{ij}(\vec{\Theta}^{in}, \hat{\Lambda})$) and those directly learned for each of the input problems ($\Theta^{out}(\Theta^{in})$). Here $\Theta$ denotes the input parameters, the $J$ and $h$ values of the input Hamiltonian in this case. The scatter plots shows that the output parameters are predicted to a high accuracy, thus illustrating the correctness of the learned response function.
\conditionaltext{\alcomm{[Do we only have $J^{in}$ in (b) or full $\Theta^{in}$? Respectively, in (c) is it $J^{out}$ only? Should be $\hat{\Lambda}$ in (c).]}}}
\label{fig:Fig1}
\end{figure*}

Whereas the input parameters can be uniquely defined as the user-specified parameters of the problem Hamiltonian, the definition of the output parameters and their relation to the Hamiltonian is not immediately obvious given that often in practice the user can only observe data samples corresponding to measurements in the computational basis at the end of the quantum evolution. The definition of the output parameters that we use in this work is based on an observation made in prior work that analog quantum computers can be seen as sampling from a thermal Gibbs distribution \cite{perdomo2016determination,nelson2022high,vuffray2022programmable,benedetti2016estimation,li2020limitations} with a Hamiltonian related to but not equal to the input Hamiltonian \cite{vuffray2022programmable}. Parameters of the output Hamiltonian associated with this Gibbs distribution extracted from data can then be used as a definition of the output parameters.

Extraction of Hamiltonian parameters from output data is related to the problem of learning of Gibbs distributions \cite{vuffray2016interaction,vuffray2019efficient,lokhov2018optimal,Klivans2017}. In an attempt to characterize the nature of the output distribution, previous works have either done partial characterization by estimating a few key parameters related to output Hamiltonian, such as effective temperature \cite{perdomo2016determination}, or have been forced to do a complete study of a few qubits only, by collecting data from a large number of runs with the same input parameters  \cite{vuffray2022programmable}. Such an approach of running specific calibration runs is not convenient because one need to block out the device for a long period of time, and is generally extremely expensive as it requires a collection of data for specific problem inputs. Ideally, we would like to be able to learn the response function on the fly for larger systems without using expensive dedicated runs and by simply observing the pairs of input parameters and output samples produced by the running analog quantum computers for independent purposes.

In this work, we propose an efficient streaming learning algorithm that addresses this challenge and learns a response function of an analog quantum computer from arbitrary pairs of input parameters and output data samples obtained from unrelated computations. Our estimator is consistent under the assumption that the correct response belongs to a proposed function family. We illustrate the efficacy of our learning algorithm on several analog quantum computing platforms based on superconducting and neutral-atom programmable arrays, gaining insight on the approximate form of the response function and the range of parameters characterizing the output distributions of these quantum devices.

\section{Results}

\subsection*{Response function learning problem}

Existing commercially available realizations of analog quantum computers primarily implement the time-dependent quantum transverse field Ising model \cite{dwave-advantage,wurtz2023aquila}. For this reason and for simplicity, we focus on this model class in presentation of our results. However, our results directly generalize to more complex model classes or measurement setups, using recent advances in learning of Gibbs distributions for general discrete models \cite{vuffray2019efficient, jayakumar2024learning}.

Here, we consider analog quantum devices described in terms of individual interacting qubits, measured in the computational basis after dynamic evolution. We associate the classical binary measurement output of the $i-$th qubit with a classical binary spin variable $\sigma_i$ taking values $+1$ or $-1$. The longitudinal part of the user-provided transverse-field Ising model Hamiltonian takes the form
\begin{equation}
    \mathcal{H}^z_{in}(\vec{\Theta}^{in}) = \sum_{i = 1}^n \sum_{j = i+1}^n J^{in}_{ij} Z_i Z_j + \sum_{i = 1}^n h_i^{in} Z_i,
\end{equation}
where $Z_i$ are Pauli operators, $\{J^{in}_{ij}\}$ are couplings between qubits, and $\{h_i^{in}\}$ are local fields. Generally, the input is given both in terms of Hamiltonian parameters that we will collectively refer to as $\vec{\Theta}^{in}$, and other hardware-specific parameters $\vec{\gamma}^{in}$ related to the parametrization of the dynamic evolution to other operational characteristics of the device. In full generality, the response function can be defined with respect to the full set of input parameters $(\vec{\Theta}^{in}, \vec{\gamma}^{in})$. However, due to the differences in specific operational parameters and constraints on the implementable modes of dynamic evolution in different analog quantum computing platforms, in this work we deliberately choose to consider response functions under fixed parameters $\vec{\gamma}^{in}$ that predict the output parameters based on $\vec{\Theta}^{in}$ exclusively. Under similar settings of the dynamic evolution and the same class of input Hamiltonians, this would allow us to directly compare different analog quantum computing platforms, as we explain below.

Under fixed device-specific parameters $\vec{\gamma}^{in}$ and for input problems characterized by the input Hamiltonian $\mathcal{H}^z_{in}(\vec{\Theta}^{in})$, we assume that computational basis measurement after the quantum evolution represent a sample $\vec{\sigma} = \{\sigma_i\}_{i = 1, \ldots, n}$ from a classical  Gibbs distribution:
\begin{equation}
    \mu_{out}(\vec{\sigma}|\vec{\Theta}^{in}) = \frac{1}{Z} \exp\left(\mathcal{H}_{out}(\vec{\sigma},\vec{\Theta}^{in})\right).
\end{equation}
In general, $\mathcal{H}_{out}$ can be a sophisticated function that is outside of the class of Ising modell. Prior work \cite{vuffray2022programmable} has shown that at least for some types of programmable analog computing devices and device characteristics $\vec{\gamma}^{in}$, higher-order terms are significantly weaker, and that $\mathcal{H}_{out}$ is well represented by an Ising model with pairwise interactions itself, although it may have additional couplings compared to the input Hamiltonian. Based on these results, we use the following anstaz for $\mathcal{H}_{out}$:    
\begin{equation}\label{eq:Eout_ising}
   \mathcal{H}_{out}(\vec{\sigma},\vec{\Theta}^{in}) = \sum_{i = 1}^n \sum_{j = i+1}^n J_{ij}^{out}(\vec{\Theta}^{in})\sigma_i \sigma_j + \sum_{i = 1}^n h_i^{out}(\vec{\Theta}^{in}) \sigma_i.
\end{equation}

In order to quantify the dependence of the output parameters on the input parameters, we need to select the responses $J_{ij}^{out}(\vec{\Theta}^{in})$ and $h_i^{out}(\vec{\Theta}^{in})$ as a member of a parametric function family. Such a formulation will allow us to formulate a learning problem of finding the parameters that best approximate the actual response. In the most generic form, we can write the output interaction coefficients as
\begin{align}\label{eq:Jout}
&J_{ij}^{out}(\vec{\Theta}^{in}) = F_{ij}( \vec{\Theta}^{in}, \vec{\Phi}_{ij} ),\\
&h_i^{out}(\vec{\Theta}^{in}) = F_{i}( \vec{\Theta}^{in}, \vec{\Phi}_{i} ) \label{eq:hout}.
\end{align}
Here, $\vec{\Phi}_{ij}$ and $\vec{\Phi}_{i}$, denoted collectively as $\vec{\Phi}$, gives the collection of parameters that fix the response function in the parametric family. In this work, we choose these to be the most general quadratic function of $\vec{\Theta}^{in}$:
\begin{align}\label{eq:response}
F_{ij}(\vec{\Theta}^{in}, \vec{\Phi}_{ij}) = c_{ij} +  ( \vec{\Theta}^{in} )^T \vec{\beta}_{ij} + ( \vec{\Theta}^{in} )^T {\chi}_{ij} \vec{\Theta}^{in},\\ 
F_{i} (\vec{\Theta}^{in}, \vec{\Phi}_{i}) = c_{i} +  ( \vec{\Theta}^{in} )^T \vec{\beta}_{i} + ( \vec{\Theta}^{in} )^T {\chi}_{i} \vec{\Theta}^{in}.
\end{align}
Here we have split $\vec{\Phi}$ in to three different components according to their physical meaning. For a total of $D$ input parameters, i.e $\vec{\Theta} \in \mathbb{R}^D$, these three components can be interpreted as follows The $c_{ij}, c_i$ parameters denote the constant offset in the output Hamiltonian independent of the input parameters.  The $\vec{\beta}_{ij},\vec{\beta}_i \in \mathbb{R}^{D}$  are the linear response parameters. Specifically, the diagonal part of the linear response ( the coefficients that affect the contribution of ${J^{in}_{ij}}$ to   ${J^{out}_{ij}}$)  measures the effective inverse temperature of the chip. If the device is an ideal Gibbs sampler for the input, then these diagonals coefficients would be the only non-zero response parameters (i.e. $ \mathcal{H}_{out}(\vec{\sigma}, \vec{\Theta}^{in}) = \beta~\mathcal{H}_{in}(\vec{\sigma}, \vec{\Theta}^{in})$ in this case). The $\vec{\chi}_{ij}, \vec{\chi}_{i} \in \mathbb{R}^{D \times D}$ are symmetric susceptibility matrices that determine the leading-order non-linear response parameters. In principle, higher order terms could be added to the response function. But we find that the quadratic form is effective enough to predict the output Hamiltonians in our experiments. Adding higher order terms will also increase the number of free parameters in the associated learning problem, potentially making the learning of the response function more data intensive.

\subsection*{The learning algorithm}
 Given a response function of the form given in \eqref{eq:response}, we want to devise an efficient algorithm that can learn this response given data from an analogue quantum device. Prior approach to this problem used a two-step approach to learn the response. First a dataset of input and output Hamiltonians were constructed by first running a suite of input jobs and learning the output Hamiltonian for each input by solving the inverse Ising problem for each of them \cite{vuffray2022programmable}. Then a linear regression was performed on this dataset to learn the response. This was a computationally intensive procedure that was limited to only $4$-qubits. The large number of optimization problems that needed to be solved limited the scalability of this approach. We improve on this prior approach by using a generalized version of graphical model learning \cite{vuffray2019efficient}, which solves a single optimization problem to then find the correct response function. Our algorithm is a data-driven approach that uses the output measurement data from different inputs (jobs run by users). We do not require any specially tailored input instances to be run on the machine. This algorithm can collect data from any input instances run by users and learn the response from this in a streaming fashion.  

Let $\Gamma$ be the set of all possible inputs that can be given to the analogue quantum device. Notice that this is always a large but discrete set in all practical devices. Further let us assume that the output distribution for any set of inputs always has an Ising form and that the response functions are quadratic. 
Now for a set of variable response parameters $\vec{\Lambda}$, define the following \emph{local energy } function for every qubit $i$ in the system,
\begin{multline}\label{eq:local_energy}
E^{(i)}_{out}(\vec{\sigma}, \vec{\Theta}^{in}, \vec{\Lambda}) = \sigma_i F_i(\vec{\Theta}^{in}, \vec{\Lambda}_i)  +  \sigma_i \sum_{j \neq i}   \sigma_j F_{ij}(\vec{\Theta}^{in},\vec{\Lambda}_{ij}) .
\end{multline}

Notice that if the parameters $\vec{\Lambda}$ in the above expression match the true system response parameters $\vec{\Phi}$  then $E^{(i)}_{out}$ gives precisely the terms in $\mathcal{H}_{out}$ that are connected to the spin $\sigma_i.$

Now in the ideal setting, given access to $\Gamma$, the response parameters that influence each qubit $i$ can be learned by minimizing the following loss function known as the \emph{Interaction screening loss} \cite{vuffray2016interaction, lokhov2018optimal, jayakumar2020learning},
\begin{equation} \label{eq:is-full}
\lc_i(\vec{\Lambda}) =  \sum_{\vec{\Theta}^{in} \in \Gamma} \mathbb{P}(\vec{\Theta}^{in})~ \underset{\vec{\sigma} \sim \mu_{out}(.|\vec{\Theta}^{in})}{\mathbb{E}} \exp \Bigg ( - E_{out}^{(u)}( \vec{\sigma}, \vec{\Theta}^{in}, \vec{\Lambda})  \Bigg)
\end{equation}
Here $\mathbb{P}(\vec{\Theta}^{in})$ is an arbitrary distribution according to which the inputs are chosen. In the \SupI~we show that for any distribution $\mathbb{P}$,  the minimum of this loss is achieved when the variables $\vec{\Lambda}$ are equal to the true system response parameters. 

In practice, we are provided with finitely many input instances and corresponding measurement shots, which allows us to approximate both the outer sum over instances and the inner expectation value over $\mu_{out}$. To this end assume that $\vec{\Theta}^{in}_1, \ldots, \vec{\Theta}^{in}_M$ are $M$ distinct input instances (sampled from some $\mathbb{P}$), to the device and let $\vec{\sigma}^{(k,1)}, \ldots, \vec{\sigma}^{(k,m)}$ be the result of $m$ measurement shots obtained from the device when programmed according to the parameters $\vec{\Theta}_k$. From this input/output dataset the interaction screening loss function can be approximated as follows,
\begin{equation} \label{eq:is-approx}
\hat{\lc}_i(\vec{\Lambda})  = \frac{1}{mM} \sum_{k = 1}^M \sum_{t = 1}^m \exp \Bigg( - E^{(u)}_{out}( \vec{\sigma}^{(k,t)}, \vec{\Theta}^{in}_k, \vec{\Lambda}) \Bigg)
\end{equation}

Now if we use a quadratic response function as in \eqref{eq:response}, then the term in the exponent is always linear in the $\vec{\Lambda}$ variables. This then implies that $\lc$ is a convex function of the $\vec{\Lambda}$ variables. Hence this loss can be easily optimized using gradient descent or its many variants.  We can also use other non-linear parameterizations of the response like neural-nets in the same loss function \cite{jayakumar2020learning}. This can lead to a richer class of response functions at the cost of making the loss function in \eqref{eq:is-approx} non-convex.

For the response parameters to be approximately recoverable from the above loss function, it is necessary for the loss function in \eqref{eq:is-approx} to have a unique global minimizer in its domain. This can be guaranteed by having a certain number of independent input instances that depend on the number of input programmable inputs to the device. For a device that has $D$ real programmable parameters ($\vec{\Theta}^{in} \in \mathbb{R}^D$) we find that $\frac{(D + 1)(D+2)}{2}$ independent instances are sufficient to uniquely fix the response function (see \SupI).

Now from the properties of this loss function, the value of $\Lambda$ that minimizes this loss function ( we call this $\hat{\Lambda})$ in the results) will be a good approximation the true response parameters.

\begin{figure*}[!t]
    \centering
    \begin{minipage}[c]{.32\textwidth}
    \subfloat[\label{fig:quera_conn}]
    {\includegraphics[height=.4\textwidth]{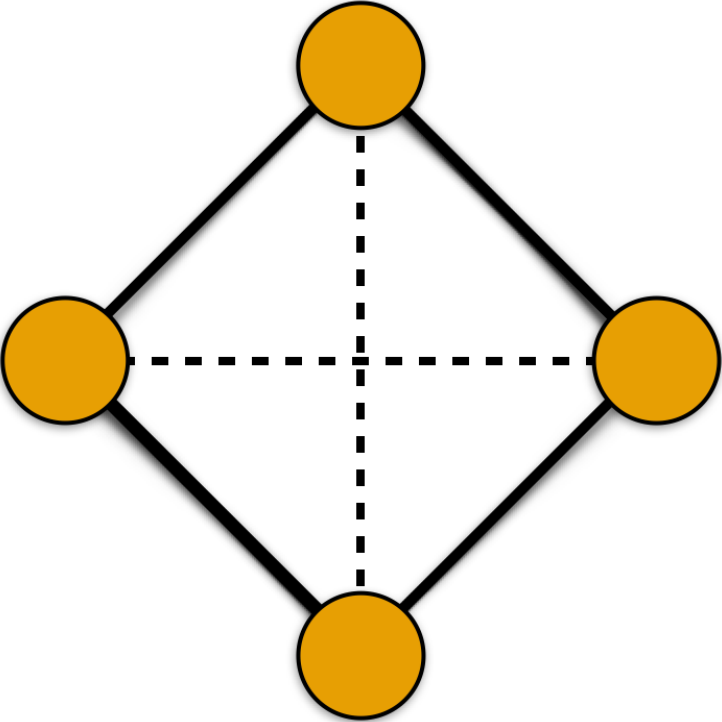}}\\
    \subfloat[\label{fig:dwave-quera}]
    {\includegraphics[width=\textwidth]{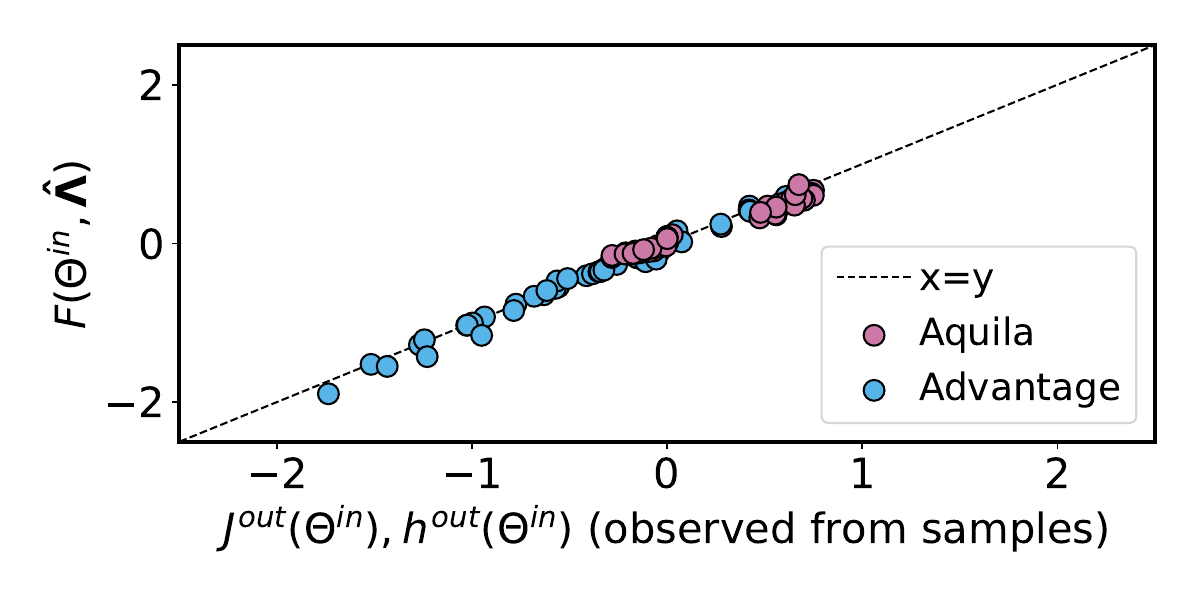}}
    \end{minipage}
    \hfill
    \begin{minipage}[c]{.32\textwidth}
    \subfloat[\label{fig:quera-atoms}]
    {\includegraphics[height=.4\textwidth]{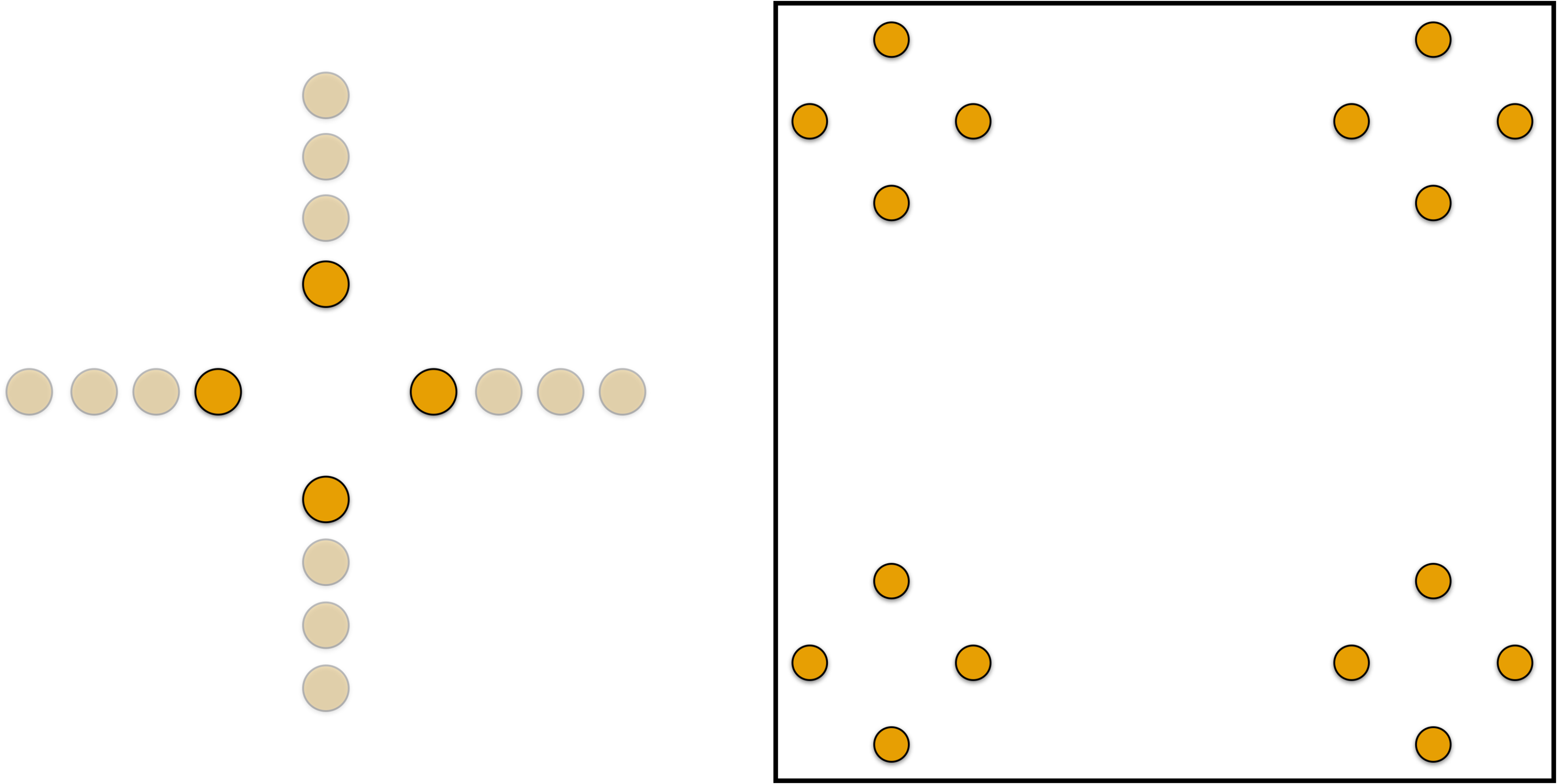}}\\
    \subfloat[\label{fig:quera-schedule}]
    {\includegraphics[width=\textwidth]{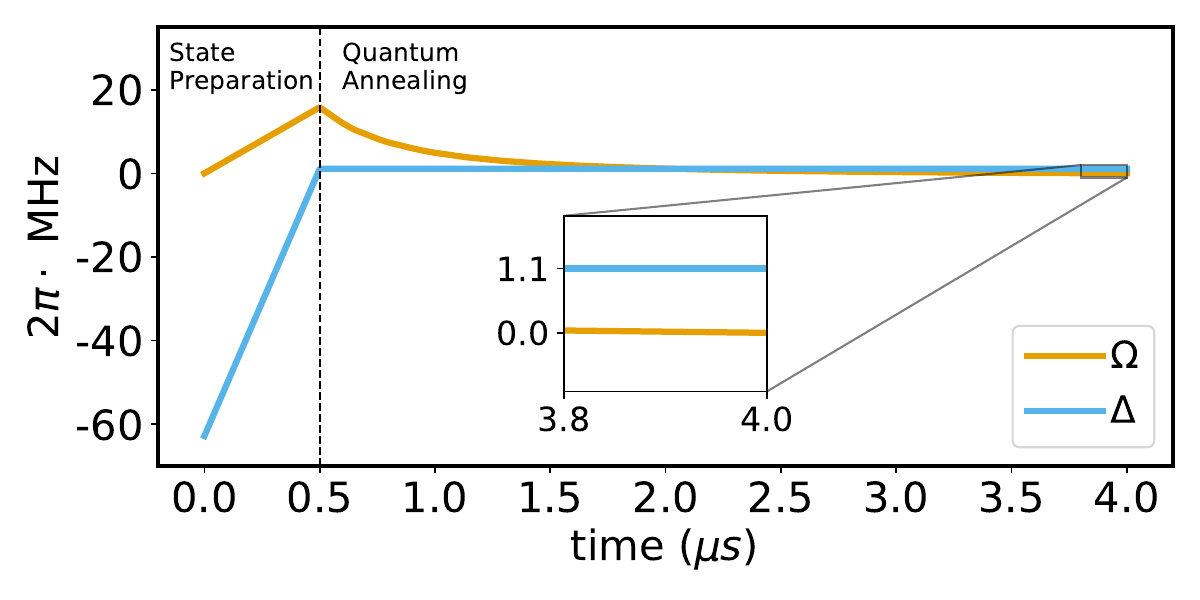}}
    \end{minipage}
    \hfill
    \begin{minipage}[c]{.32\textwidth}
    \subfloat[\label{fig:dwave-tile}]
    {\includegraphics[height=.4\textwidth]{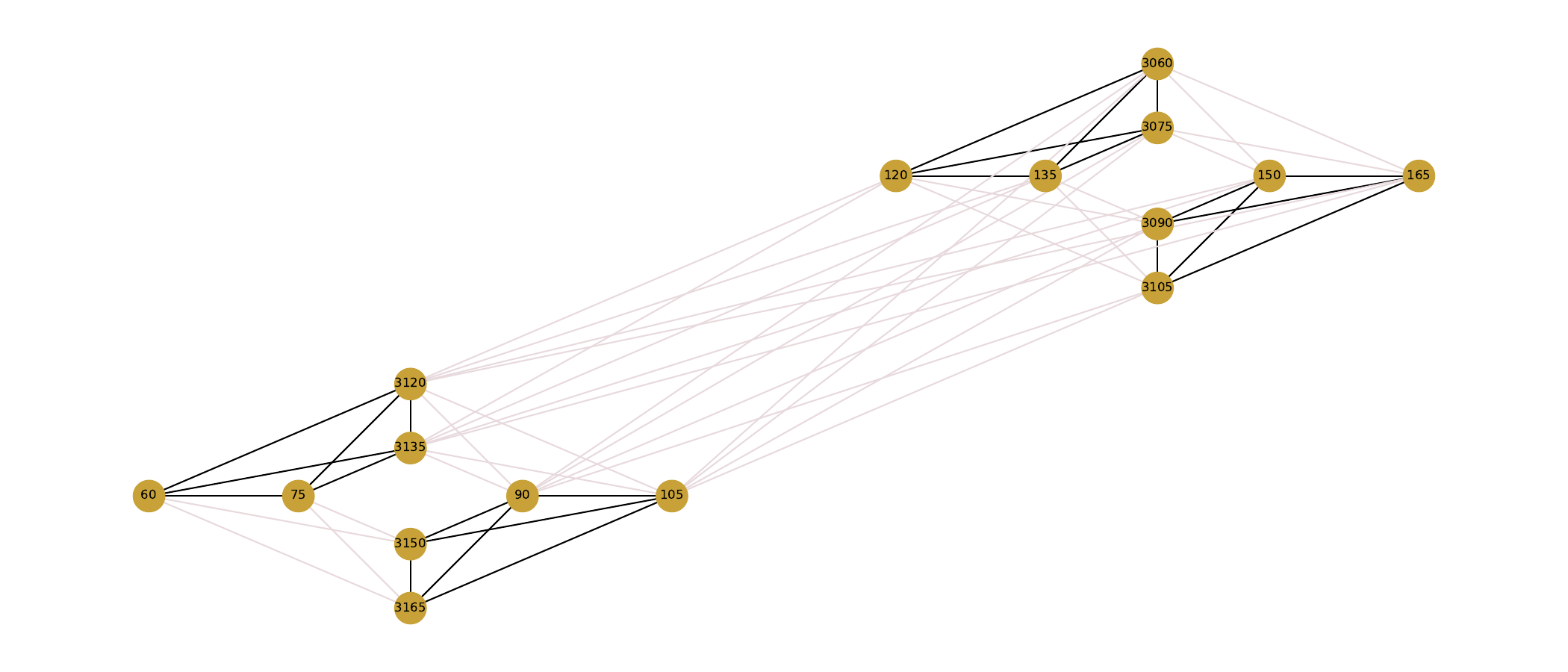}} \\
    \subfloat[\label{fig:dwave-schedule}]
    {\includegraphics[width=\textwidth]{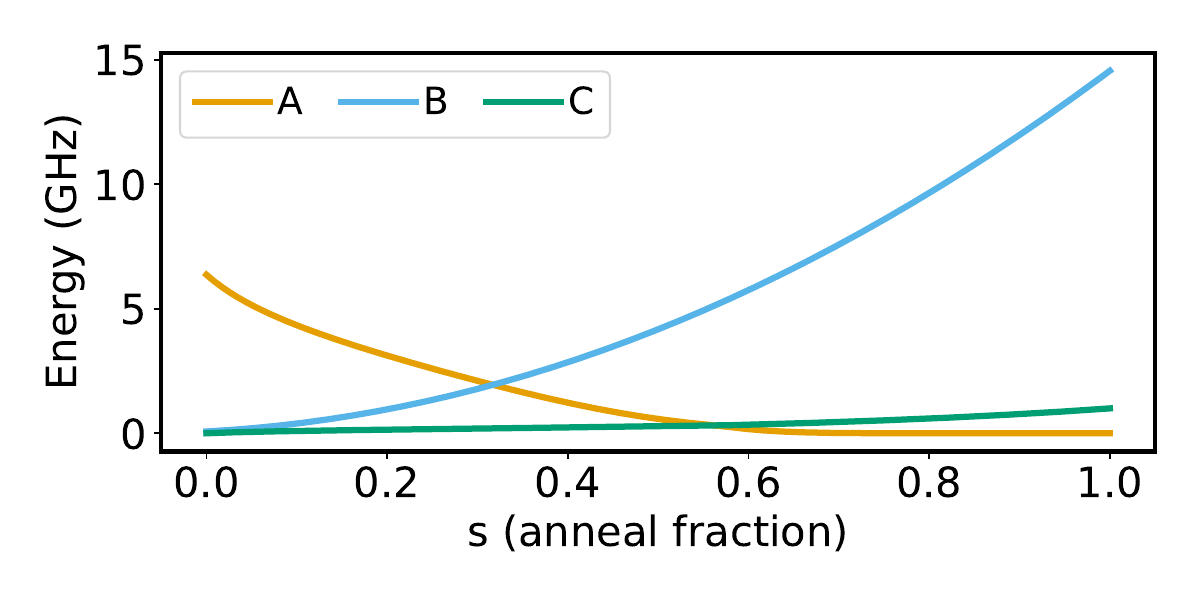}}
    \end{minipage}
    \caption{{\bf Comparison of learning averaged response function on neutral-atom and superconducting programmable arrays.} (a) The connectivity of the four qubit model. The dashed lines are used to highlight the \textit{weaker} couplings with respect to the solid lines. (b) Reconstruction results using samples from Aquila and Advantage platforms. (c) An illustration of atom positions used on the Aquila device. Atom positions are sampled randomly as described in Appendix~\ref{app:quera-atom} and tiled to the corners of the computing platform to increase total number of samples. (d) The annealing schedule that is used for the Aquila experiments. More details are provided in Sec.~\ref{sec:quera-setup}. (e) A part of the Advantage chip showing four different Hamiltonian instances tiled. The both grey and black links exist in the chip but the grey links are set to zero coupling in the input. Tiling different instances across the chip in this fashion allows us to average out qubit specific effects and results in a better comparison with the Aquila device.  (f) The annealing schedule that is used for the Advantage experiments. More details are provided in Sec.~\ref{sec:dwave-setup}.}
    \label{fig:recons}
\end{figure*}

The convex loss function in \eqref{eq:is-approx} is also amenable to optimization in an online fashion \cite{hazan2022introduction}. This allows us to update our estimate of our response parameters in the background using information from jobs being run by users on the machine. This update can be performed using the online variant of gradient descent\cite{hazan2007adaptive}. In this paper we do not use the online approach as it requires a more fine-grained access to the quantum hardware. However, the online variant would be of use to hardware maintainers who can continuously monitor the  Hamiltonian instances being run on a chip. This stream of  data can then be used to update and refine the response function associated with the hardware.

\subsection*{Experimental setup}

To test the performance of our response learning method, we conducted several experiments on two different analog quantum computing platforms, namely D-Wave quantum annealers $(\emph{Advantage},~ $2000Q$)$ \cite{dwave-advantage} and QuEra's neutral atom quantum computer  (\emph{Aquila)} \cite{wurtz2023aquila}. The response learning method will work for any analog system that produces a strictly positive output distribution. This is because, such distributions can always be represented as the Gibbs distribution of an output Hamiltonian at finite temperature, and the response function learning approach can be used to learn the mapping between the input parameters and the output Hamiltonian. In our experiments and for comparison purposes, we perform tests on quantum annealing protocols since all the hardware systems mentioned above have the native capacity to implement some type of quantum annealing (however, our approach applies to arbitrary dynamic protocols).

We perform three main experiments to show the scalability and the versatility of our approach. In all experiments a set of Hamiltonian parameters ($\vec{\Theta}^{in}$) are provided to the platforms.  The input parameters are randomly sampled from a uniform distribution as described in \emph{Methods}, Sec.~\ref{sec:methods}. The computers are then programmed to perform an annealing schedule to prepare a low energy state of the input Hamiltonian $\mathcal{H}^z_{in}(\vec{\Theta}^{in})$. The output samples are collected by performing a hardware  native (computational basis)  measurement at the end of the anneal. The annealing schedules for two types of platforms are provided in Fig.~\ref{fig:quera-schedule} and \ref{fig:dwave-schedule}.

Using the procedure described above, we construct a dataset that consists of different instances of input Hamiltonians along with the corresponding samples (i.e., measurement shots) produced by the quantum anneal. We use a part of this dataset to evaluate and optimize the objective function described in \eqref{eq:is-approx} for each qubit in the Hamiltonian. The optimal response function hence found  by optimizing $\hat{\lc}_i$ gives us a prediction for all the couplings of the output Hamiltonians connected to qubit $i$. The remaining part of the dataset, which contains Hamiltonian instances that were not used to train the response function, is used to test the accuracy of the learned response function for predicting the output.

The primary hypothesis we aim to test in these experiments is whether the Hamiltonian predicted by the learned response function closely aligns with the actual output Hamiltonian, which models the devices' output distribution. To test this, we need to have an independent estimate of the output Hamiltonians produced by the input instances in the test set. To this end we use the \textit{RISE} algorithm \cite{lokhov2018optimal,vuffray2016interaction,vuffray2022programmable}  to learn the output Hamiltonians from the samples in the test dataset. RISE is an efficient algorithm to solve the inverse Ising problem \cite{chayes1984inverse}, i.e., given samples from a Gibbs distribution on binary variables, RISE can efficiently reconstruct the Hamiltonian of the Gibbs distribution. The response function learning procedure that we outlined in the previous section can be seen as a generalization of RISE. The crucial difference here being that RISE produces the output Hamiltonian for a single instance and does not produce a response function that can predict the output Hamiltonians associated with new, unseen problem instances. In contrast, the learned response function can predict the output Hamiltonian from just the input parameters. It does not require access to the samples produced from the hardware to make this prediction.

In all results, we report the inferred Hamiltonian parameters ($\vec{\Theta}^{out}$) of unseen inputs, using the learned response parameters ($\hat{\Lambda}$). This is compared against the output energy function given by the response function. In these tests, we limit ourselves to relatively small systems in order to allow for comparisons involving explicit reconstructions with RISE. To validate the statistical significance of the results, we resample from the Hamiltonian that is learned using the RISE algorithm and re-learn the model using the new samples. We found that the variation of the learned parameters were not significant enough to be plotted.

\subsection*{Experiment 1: Learning single-cell response function on a superconducting platform}
In the first experiment, we focus on a specific part of the chip on D-Wave 2000Q platform  as illustrated in Fig.~\ref{fig:Fig1}. The experimental setup is similar to the one used in  Reference \cite{vuffray2022programmable}, which was used to learn response functions for four qubit systems. The response learning method in \cite{vuffray2022programmable} used RISE to first learn the output Hamiltonians and linear regression to then find a response. The scalability of this two-step approach was limited. With the single step learning algorithm presented here, we illustrate our approach on a 16-qubit system.

\conditionaltext{\alcomm{[Add this comment on terms kept in the response.]}}
\conditionaltext{
\ajcomm{Brief comments on which terms were kept in the response. That gradient descent was used to promote scalability}}

For our tests, we learned the response function using  $M= 225$ input Hamiltonians on  16 qubit programs, where each Hamiltonian was used to produce $4 \times 10^6$ output samples. We plot the inferred Hamiltonian parameters, using the learned response function on one unseen instance against the one obtained using \textit{RISE} in Fig.~\ref{fig:Fig1}. The trend along the $x=y$ line shows near-perfect agreement. This experiment thus illustrates the validity of the novel streaming loss function introduced in this work \eqref{eq:is-full}-\eqref{eq:is-approx}.

\subsection*{\label{sec:quera-dwave}Experiment 2: Learning averaged response function on neutral-atom and superconducting programmable arrays}

Next we use the response learning algorithm on two different types of hardware; superconducting (Advantage by D-Wave~\cite{dwave-advantage}) and neutral atom (Aquila by QuEra~\cite{wurtz2023aquila}). Neutral atom platforms have significantly different working principles than superconducting hardware. Neutral atom devices use Rydberg atoms as qubits and interactions between qubits happen through the van der Waals force, whose strength is inversely proportional to the distance between atoms as $\sim (1/|\vec{r}_i-\vec{r}_j|^6)$, where $\vec{r}_i$ is the position of the $i^{th}$ atom. None the less, the Hamiltonian of these systems can be shown to be equivalent to an Ising model Hamiltonian with additional single body terms.   

To understand the differences in these two types of systems, we would like to run the same type of Hamiltonians on both hardware. For this purpose, we rewrite 
the neutral atom Hamiltonian in the Pauli basis, taking into account the global control (see Section~\ref{sec:quera-setup} for details),

\begin{align}
\frac{\mathcal{H}(t)}{\hbar} &= \frac{\Omega(t)}{2} \sum_j X_j + \sum_{j =2}^n \sum_{k=1}^{j-1} \frac{V_{jk}}{4} Z_j Z_k  \\ \nonumber
& - \frac14\sum_{j=1}^{n} \sum_{k\neq j} V_{jk} Z_j   +\frac{\Delta(t)}{2} \sum_{j} Z_j.
\label{eq:quera-ham-pauli}
\end{align}

The Hamiltonian in the Pauli basis highlights an important facet of the neutral atom systems, which the the coupling term ($V_{jk}$) appearing in front of the field terms ($Z_j$). This results in a bias on the field terms that can not be completely isolated, since there is no local control of these terms available at the time of writing this paper~\cite{wurtz2023aquila}. We take this into account when designing the the schedule (see Appendix~\ref{app:schedule}), which is illustrated in Fig.~\ref{fig:quera-schedule}.

As opposed the superconducting systems, neutral atom platforms do not have unique physical qubits. This makes them inherently different from superconducting quantum computers, which have physical qubits located on a specific part of a chip. For this reason, while collecting data from the superconducting device, we also tile the chip with the same instance while sampling the data, in order to remove the noise effects that are dedicated to specific qubits. Note that the tiling is only done for this experiment: here, we learn and compare \emph{average} responses for different qubit cells as opposed to \emph{single-cell} responses. This tiling also helps to increase the effective shot rate of the neutral atom device, whose native shot rate is slower that superconducting devices.

After bringing two technologies to the same footing, we perform four qubit experiments that use 100 and 295 training instances and 10k and 1M samples respectively on the neutral atom and the superconducting hardware.  We collect fewer instances from the neural atom device as we observe that the response function with just linear terms are sufficient to model the output Hamiltonians in this system. Since the input Hamiltonian has $D = 10$ parameters, $D+1 = 11$ and $(D+1)(D+2)/2 = 66$  independent instances are sufficient here to fix the response for respectively linear and quadratic response functions. Empirically we observe that more instances improve the quality of the learned response function in the testing phase.

We perform testing on the same five unseen instances to compare their behaviour and plot it against the result obtained using RISE in Fig.~\ref{fig:dwave-quera}.

The distribution of the response functions are visibly different in two cases. We observe a separation in terms of the response function in the neutral atom case. The coupling terms ($J_{ij}^{out}$) are centered around zero, while the field terms ($h_{i}^{out}$) are shifted with a positive bias. \conditionaltext{\ctcomm{Should we talk about why we think this is the case here.}} On the other hand, the output of the superconducting hardware shows a more homogeneous distribution. Response of the superconducting hardware span larger values, which points to a difference in the strength interactions in the two devices. While it seems like the neutral atom system has weaker interactions, this is a feature of the specific instances that we chose to run on the system. Stronger interactions could have been observed had we reduced the distance between the atoms in the device. But then more samples from each input Hamiltonian would be required to learn the response funciton in this stronger coupling regime. This is due to the lower bounds on the inverse Ising problem which state that the sample complexity of this learning task should scale exponentially with the strength of interactions \cite{santhanam2012information}. Due to the considerably low shot rate of the neutral atom devices ($<$ 10 shots per second \cite{wurtz2023aquila}), currently it is not feasible to benchmark these devices using these techniques in the stronger coupling regimes. On the other hand, the superconducting device has a much higher shot rate (observed around 5000 shots per second) and hence input Hamiltonians with stronger couplings can be used in this case. Overall, the efficacy of our method persists in both cases as demonstrated by the alignment along the diagonal line.  The response function coefficients learned for these experiments can be found in Appendix C (\figurename 8 and \figurename 9).

\subsection*{Experiment 3: Learning single-cell response function on two generations of superconducting platforms}

\begin{figure*}[!t]
    \centering
    \begin{minipage}[c]{.32\textwidth}
    \subfloat[\label{fig:4q_conn}]
    {\includegraphics[height=.4\textwidth]{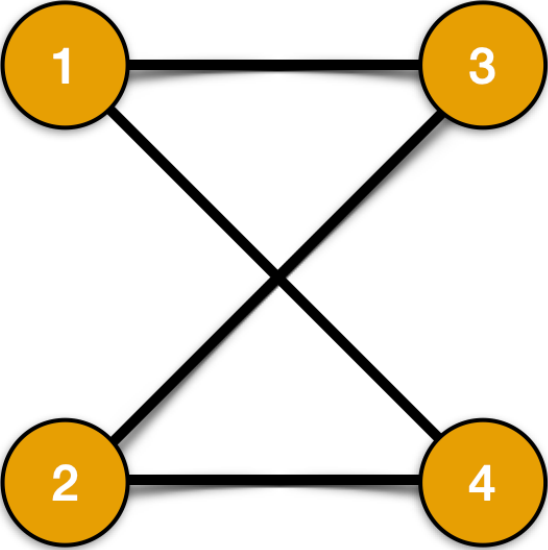}} \\
    \subfloat[\label{fig:dwave}]
    {\includegraphics[width=\textwidth]{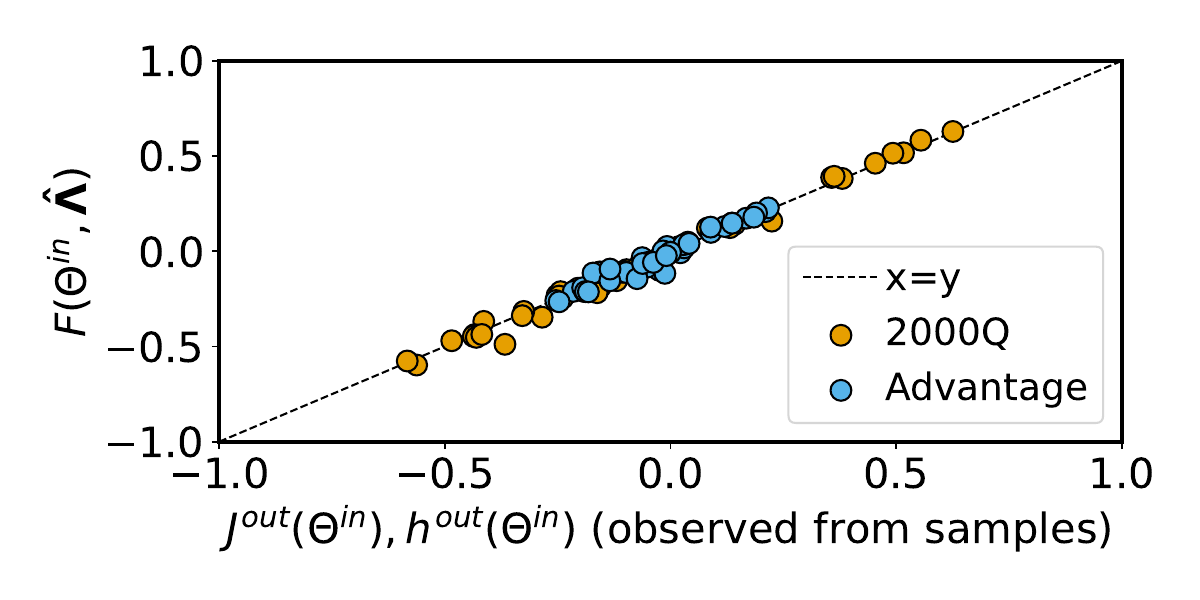}}
    \end{minipage}%
    \hfill%
    \begin{minipage}[c]{.66\textwidth}
    \subfloat[\label{fig:dwave-terms}]
    {\includegraphics[width=\textwidth]{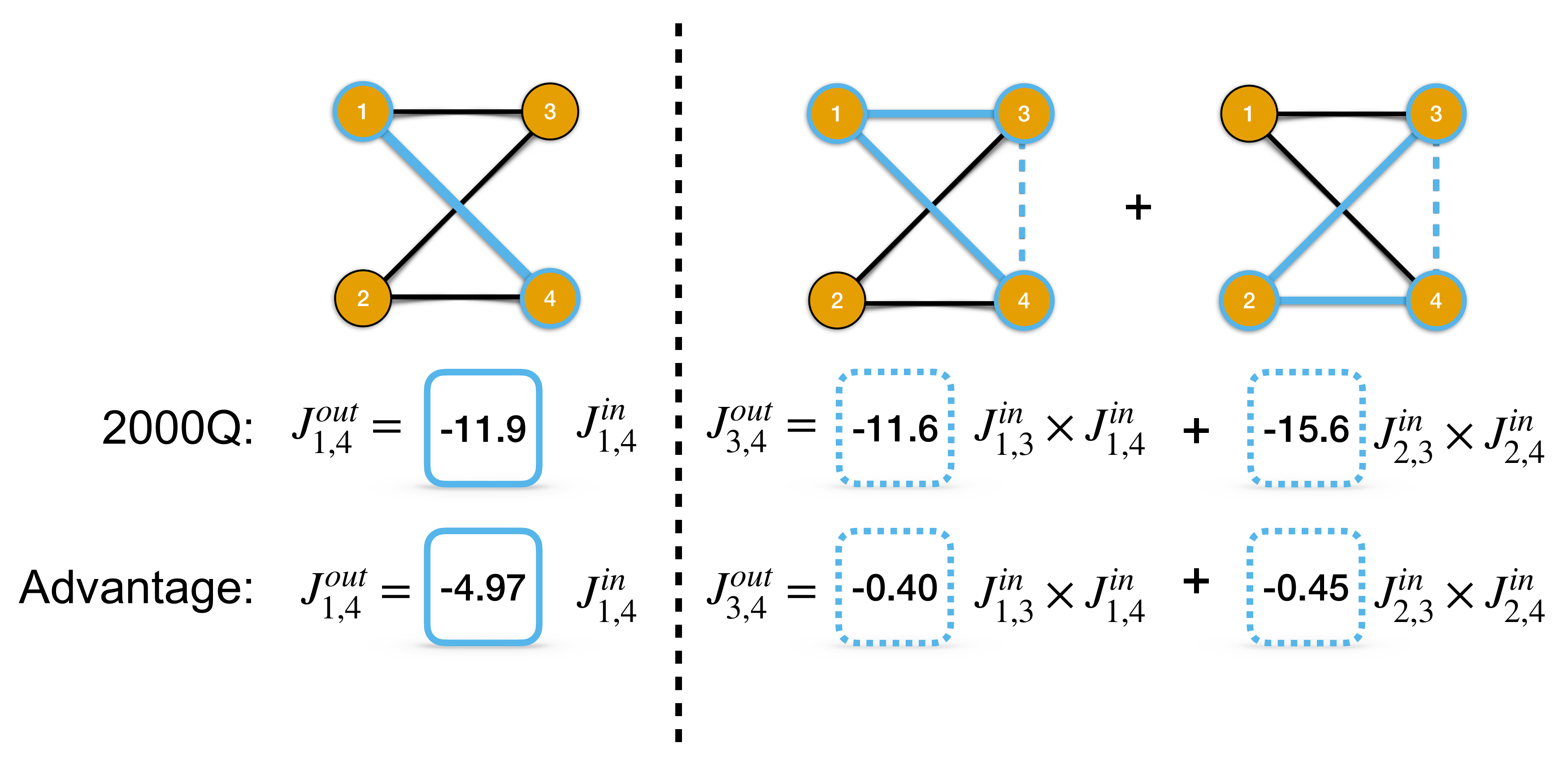}}
    \end{minipage}
    \caption{{\bf Term by term comparison of the response functions obtained on D-Wave 2000Q and Advantage.} This illustration presents two terms of the response function. The first term $J_{1,4}^{out}$ is chosen to illustrate the difference in the effective temperatures of the two devices. The second term $J_{3,4}^{out}$ is a non-existent connection in the input couplings, although it makes an appearance in the final reconstruction. For the second term we only present the most significant terms, refraining from showing the rest to simplify the drawing (full details are provided in the Supplementary Material Appendices).}
    \label{fig:dwave-4q}
\end{figure*}

 In our final experiment, we compare two superconducting hardware; D-Wave 2000Q (old generation) and Advantage (new generation). The aim of this experiment is to demonstrate the the utility of the response coefficients as a benchmarking tool.   We use the same 245  input instances on both devices to train with 4M and 100k samples respectively. The inferred response of five unseen instances are plotted against the results of the \textit{RISE} method in Fig.~\ref{fig:dwave}. For these experiments, we use a $4$-qubit input Hamiltonian with two inter-qubit connections set to zero. The aim to this configuration is to then study the value of this coupling in the output Hamiltonian. The value of this spurious coupling in the output will act as a proxy for the noise in the hardware

Following the success of the other experiments, we once again show that our method is able to match the outcome of RISE method with great precision. We observe a discrepancy in response functions of the two devices as evidenced by the span of the response function values.

It is possible to better understand the difference between two devices by looking at the learned parameters of their response function. In Fig.~\ref{fig:dwave-terms} we illustrate some terms of the response function. Comparison of the $J_{1,4}^{out}$ term highlights the decrease in effective inverse temperature in between two generations. Next, we compare the $J_{3,4}^{out}$ term, which the spurious  term that appears as these two qubits are not connected in the input Hamiltonian. The magnitude of the coefficients indicate the significant drop in spurious links between qubits in the Advantage device. The values we report for 2000Q match the previous study of Vuffray et al.~\cite{vuffray2022programmable}, where these terms are initially reported. Notice that the sign of the linear terms are different from what is reported in \cite{vuffray2022programmable}. This is  matter of convention. Here we have chosen to define the output distribution such that the maximum of the output energy function approximately corresponds to the ground state of the input Hamiltonian.

\section{Discussion}
In this work, we introduce a new, scalable approach to learning the response functions for analog quantum computers.
We have demonstrated that our algorithm on different AQC platforms to learn the response  based solely on the input Hamiltonian parameters. This approach can be extended to learn the response to a more diverse set of input parameters, including general quantum dynamics beyond quantum annealing.

From our annealing tests, we found that a quadratic response function is sufficient to predict the output Hamiltonian for both superconducting and neutral atom platforms. This simple form of the response opens up the possibility of using the learned response parameters as a benchmarking tool. This is demonstrated in Fig. 3, where we see improvements in the D-Wave hardware reflected in the response coefficients.

Given better access to the AQC device our algorithm can be deployed in an  online fashion by hardware developers to train a surrogate model for larger parts of the chip or the entire chip itself. The computational cost of this task can be further improved by including prior information, e.g. restricting the reconstruction to relevant quadratic terms only (e.g., those presented in Fig. 3). Another approach, which is not explored in this work, would be patch together response functions of smaller portions of the chip to construct a model for the whole chip. The practical trade-offs introduced by such an approach is unknown and will be studied in future works. 

An important downstream application of such a response function will be for error mitigation. As the response function correctly predicts the output Hamiltonian, applying the inverse of this function on a target Hamiltonian will tell us what model parameters the user should input so as to mitigate the noise in the device and produce more accurate samples from the desired target Hamiltonian. A thorough examination of this application is beyond the scope of this paper, but will be demonstrated in an upcoming work.

\section{\label{sec:methods} Methods}

\subsection{\label{sec:dwave-setup} Setting up the superconducting platform experiments}

On superconducting devices, the time dependent control of the system can be defined as follows,

\begin{align}
     \mathcal{{H}}(s) &= -\frac{A(s)}{2} \sum_j X_j  \\ \nonumber
     &+\frac{B(s)}{2} \left( \sum_{j =1}^n \left( \sum_{j>k} J_{jk} Z_j Z_k \right) + h_j Z_j \right).
\label{eq:dwave-schedule}   
\end{align}

Quantum annealing procedure begins by setting the term $B(s=0)$ to zero and $A(s=0)$ to a non-zero value. This corresponds to a system, whose ground state is the equal superposition state in the computational basis. This state is easy to prepare and can be achieved on analog quantum computers. Then, the $A(s)$ term is gradually evolved to be zero, while the term $B(s)$ is increased to match the target final Hamiltonian. Then, if the evolution is performed slow enough, the final state will get close  to the ground state of the Hamiltonian at the end of the process, according to the adiabatic theorem~\cite{adiabatic_theorem}. We present the choice of schedule for the superconducting experiments in Fig.~\ref{fig:dwave-schedule}. We refer the reader to Ref.~\cite{dwave-annealing-schedule} for more details.

The input instances consist of a collection of Hamiltonians that are randomly sampled following Ref.~\cite{vuffray2022programmable}. To obtain these Hamiltonians, we first fix a region of the chip with the connectivity defined by the graph $G(V,E)$, where $V$ denotes the qubits and $E$ denotes the physical connections that can be programmed. Then, the field and coupling terms are sampled according to the chosen graph $G$ from the uniform distribution described as follows,

\begin{align}
     &h_{i}^{in} \in \{-0.05 : 0.01 : 0.05 \} \, \forall i \in N \\
    &J_{ij}^{in} \in \{-0.05 : 0.01 : 0.05\} \, \forall i,j \in E   
\end{align}

\subsection{\label{sec:quera-setup} Setting up the neutral-atom platform experiments}

The time evolution evolution of a neutral atom system is controlled by the folllowing time dependent Hamiltonian~\cite{wurtz2023aquila},

\begin{align}
\frac{\mathcal{H}(t)}{\hbar} &= \sum_j \frac{\Omega_j(t)}{2} \left( e^{i \phi_j(t) } | g_j \rangle  \langle r_j | + e^{-i \phi_j(t) } | r_j \rangle  \langle g_j | \right) \\  \nonumber
&- \sum_j \Delta_j(t) \hat{n}_j + \sum_{j < k} V_{jk} \hat{n}_j \hat{n}_k. 
\label{eq:quera-ham}
\end{align}

Here, $\ket{g_j}$ denotes the ground state and $\ket{r_j}$ denotes the Rydberg state on qubit $j$. The Hamiltonian has three main time dependent control parameters $\Omega_j(t)$ (Rabi frequency), $\phi_j(t)$ (laser phase) and $\Delta_j(t)$ (detuning of the driving laser field). The term $V_{jk} = \frac{C_6}{|\vec{r}_j-\vec{r}_k|^6}$ describes the strength of the van der Waals interaction, with the constant $C_6 = 862690 \times 2\pi~\mbox{MHz}~\mu \mbox{m}^6$. Finally the number operator is defined as, $\hat{n}_j = \ket{r_j}\bra{r_j}$.

Currently available neutral atom platforms only offer global control of the system~\cite{wurtz2023aquila, henriet_quantum_2020}. This means the parameters will simply be the same for all qubits, \textit{i.e.} $\Omega_j(t)= \Omega(t)$, $\Delta_j(t)=\Delta(t)$ and $\phi_j(t)=\phi(t)$. 

Similar to the experiments we performed on superconducting platforms, we need to construct a set of input instances that are obtained through perturbing the coefficients of the field and coupling terms. We achieve the same strategy by perturbing the positions of the atoms in the case of the neutral atom hardware. Four atoms are placed on the vertices of a square with the width of $10.6 \mu m$. This constitutes what we call as the \textit{base}-Hamiltonian. Then, each atom are randomly shifted on their axis to vary the coupling terms. This way the \textit{base}-Hamiltonian is perturbed and many random instances can be obtained to be used as the dataset. These four atoms are tiled to the corners of the computing plane\conditionaltext{\ctcomm{Do you have a better wording idea for this phrase?}} in order to maximize the amount of samples from each run. An illustration of this procedure can be seen in Fig.~\ref{fig:quera-atoms} and more details are provided in Appendix~\ref{app:quera-atom}.

\subsection{Details on the training procedure}

The optimization of the response function is achieved by minimizing the interaction screening loss function in (\ref{eq:is-full}). This convex loss function can be easily minimized using variants of gradient descent methods or other linear solvers. In this work, we employed the entropic descent algorithm~\cite{beck2003mirror} and IPOPT \cite{wachter2006implementation} and observed similar performance. 

Furthermore, we obtained significant improvements in total computation time and memory usage when we incorporated inductive biases from previous work~\cite{vuffray2022programmable} that allowed us to omit many terms from the final representation of the response function. This deemed very useful especially for the 16 qubit experiments, when we showed the scalability of our approach. Namely, relying on the results of Ref.~\cite{vuffray2022programmable} we were able to omit quadratic terms of the connections that are not physically present. For a more concrete example,  consider the four qubit example presented in Fig.~\ref{fig:dwave-4q}. The non-physical output coupling of the term $J_{3,4}^{out}$ contains contributions only from the physical couplings. Therefore, any parameter that would be multiplied with a non-physical coupling can be completely omitted. If not omitted, these parameters are learned as zero, therefore they do not contribute to the learning results.


\section*{Acknowledgements}
The authors acknowledge support from the Laboratory Directed Research and Development program of Los Alamos National Laboratory (LANL) under Projects No. 20240032DR and No. 20230338ER, and from the U.S. Department of Energy/Office of Science Advanced Scientific Computing Research Program. CT was supported in part by the Helmholtz Association -``Innopool Project Variational Quantum Computer Simulations (VQCS)'', and by the Quantum Computing Summer School Fellowship at LANL. This paper used computational resources provided by the Quantum Cloud Access Project and by the Darwin testbed at LANL.

\bibliography{biblio}
\bibliographystyle{plain}
\onecolumngrid

\newpage
\onecolumngrid
\newpage
\appendix
\section{Technical details on the streaming interaction screening estimator}
The fact that the minimizer of the interaction screening loss is the true response parameters can be verified by evaluating the gradient of the loss at $\mathbf{\Lambda}.$
\begin{align}
\nabla \lc_i(\mathbf{\Lambda}) &=  \sum_{\vec{\gamma} \in \Gamma} \mathbb{P}(\vec{\gamma}) \underset{\vec{\sigma} \sim \mu_{out}(.|\vec{\gamma})}{\mathbb{E}} \exp \Bigg ( - E_{out}^{(i)}( \vec{\sigma}, \vec{\gamma}, \mathbf{\Lambda})  \Bigg) \nabla E_{out}^{(i)}( \vec{\sigma}, \vec{\gamma},\mathbf{\Lambda}),\\
&=   \sum_{\vec{\gamma} \in \Gamma} \mathbb{P}(\vec{\gamma}) \frac{1}{Z(\vec{\gamma})} \sum_{\vec{\sigma}} \exp \Bigg ( E_{out}(\vec{\sigma},\vec{\gamma})- E_{out}^{(i)}( \vec{\sigma}, \vec{\gamma}, \mathbf{\Lambda})  \Bigg) \nabla E_{out}^{(i)}( \vec{\sigma}, \vec{\gamma},\mathbf{\Lambda}).
\end{align}
Now consider the function $E_{out}(\vec{\sigma},\vec{\gamma})- E_{out}^{(i)}( \vec{\sigma}, \vec{\gamma},\mathbf{\Lambda})$. If we assume that the output distribution and the responses of the device are given by \eqref{eq:Eout_ising}, \eqref{eq:Jout}, and \eqref{eq:hout} then this function will have no dependence on the value taken by $i-$the qubit, $\sigma_i$. At the same time, the derivative of $E_{out}^{(i)}$ w.r.t to any parameter in the problem is a linear function of $\sigma_i$ with no intercept. This can be plainly seen from \eqref{eq:local_energy}. Now let $\vec{\sigma}_{\setminus i}$ be the measurement outcomes of all qubits except $i$, then for some function, $G$, the gradient of the interaction screening loss can be written in the form,
\begin{equation}
\nabla \lc_i(\mathbf{\Lambda})  = \frac{1}{|\Gamma|} \sum_{\vec{\gamma} \in \Gamma} \frac{ \mathbb{P}(\vec{\gamma})}{Z(\vec{\gamma})} \sum_{\sigma_i = \pm 1} \sigma_i \sum_{\vec{\sigma}_{\setminus u}} G( \vec{\sigma}_{\setminus u}, \vec{\gamma}, \mathbf{\Lambda}). 
\end{equation}
Clearly, this vanishes at the $\mathbf{\Lambda}$ that models the output energy function. The local energy is a linear function of the variables $\mathbf{\Lambda}$. This makes the exponential of the local energy a convex function in $\mathbf{\Lambda}$ when the other variables are fixed. Now the interaction screening loss is a convex combination of such exponential functions. Since convexity of a functions are preserved under convex combinations, this implies that interaction screening loss is also convex in $\mathbf{\Lambda}$. And by known properties of convex functions, any point in its domain where the gradient vanishes must be a global minimizier. Now if we assume $|\Gamma| \geq (D+1)(D+2)/2$, then by the arguments in Section \ref{subsec:a} this will also be the unique global minimum.

\subsection{Number of instances to uniquely fix the response}\label{subsec:a}

Since the loss only depends on the local energy function, we must have enough instances such that the local energy can be uniquely identified from all possible linear functions on $X$ and $Y.$ 
\begin{align}
F_{ij}(\vec{\Theta}^{in}, \vec{\Phi}_{ij}) &= c_{ij} +  ( \vec{\Theta}^{in} )^T \vec{\beta}_{ij} + ( \vec{\Theta}^{in} )^T {\chi}_{ij} \vec{\Theta}^{in},\\ 
F_{i} (\vec{\Theta}^{in}, \vec{\Phi}_{i}) &= c_{i} +  ( \vec{\Theta}^{in} )^T \vec{\beta}_{i} + ( \vec{\Theta}^{in} )^T {\chi}_{i} \vec{\Theta}^{in}. \\
\vec{\Theta}^{in} &= \{J_{12},h_1, \ldots \}
\end{align}

Now if the input parameters form a $D-$dimensional vector then the these functions are quadratic forms in $\mathbb{R}^D.$ The total number of parameters in such a function is $(D+1)(D+2)/2$. Hence we need data from  $(D+1)(D+2)/2$ instances to uniquely fix the function. Similarly if there was no quadratic response, only $D+1$ instances would suffice.

\section{Technical details of hardware experiments}

\subsection{Hamiltonian derivation}
In this section we present details for the derivation of the neutral atom Hamiltonian. 
A quantum system evolves in time following the Schr\"odinger's equation as,

\begin{equation}
    i \hbar \dfrac{\partial}{\partial t} | \psi \rangle = \hat{\mathcal{H}}(t) | \psi \rangle,
\end{equation}

where $\hat{\mathcal{H}}(t)$ is the time dependent Hamiltonian. Let us recall the general time dependent Hamiltonian that describes a Rydberg atom array~\cite{wurtz2023aquila},

\begin{align}
\frac{\mathcal{H}(t)}{\hbar} &= \sum_j \frac{\Omega_j(t)}{2} \left( e^{i \phi_j(t) } | g_j \rangle  \langle r_j | + e^{-i \phi_j(t) } | r_j \rangle  \langle g_j | \right) - \sum_j \Delta_j(t) \hat{n}_j + \sum_{j < k} V_{jk} \hat{n}_j \hat{n}_k. 
\end{align}

Here, $\ket{g_j}$ denotes the ground state and $\ket{r_j}$ denotes the Rydberg state on qubit $j$. The Hamiltonian has three main time dependent control parameters $\Omega_j(t)$ (Rabi frequency), $\phi_j(t)$ (laser phase) and $\Delta_j(t)$ (detuning of the driving laser field). The term $V_{jk} = \frac{C_6}{|\vec{r}_j-\vec{r}_k|^6}$ describes the strength of the van der Waals interaction, with the constant $C_6 = 862690 \times 2\pi~\mbox{MHz}~\mu \mbox{m}^6$. Finally the number operator is defined as, $\hat{n}_j = \ket{r_j}\bra{r_j}$.

We are interested in expressing this equation in the Pauli basis to be able to make comparisons with the superconducting platforms. For this purpose, if we take $\ket{r_j}$ to be $\ket{1}$, then $\hat{n}_j = \frac{I - Z_j}{2}$ and $\ket{g_j}\bra{r_j} = \ket{0_j}\bra{1_j} = \frac{X_j + i Y_j}{2}$. Then, the Hamiltonian in the Pauli basis reads,

\begin{align}
\frac{\mathcal{H}(t)}{\hbar} &= \sum_j \frac{\Omega_j(t)}{2} \left( \cos( \phi_j(t)) X_j - \sin( \phi_j(t) ) Y_j \right) + \sum_{j =2}^n \sum_{k=1}^{j-1} \frac{V_{jk}}{4}  (Z_j Z_k - Z_j - Z_k) + \frac{1}{2} \sum_{j} \Delta_j(t) Z_j  \nonumber \\
&= \sum_j \frac{\Omega_j(t)}{2} \left( \cos( \phi_j(t)) X_j - \sin( \phi_j(t) ) Y_j \right) + \sum_{j =2}^n \sum_{k=1}^{j-1} \frac{V_{jk}}{4} Z_j Z_k - \frac14\sum_{j=1}^{n} \sum_{k\neq j} V_{jk} Z_j  + \frac{1}{2} \sum_{j} \Delta_j(t) Z_j.
\end{align}

Currently available neutral atom platforms only offer global control of the system~\cite{wurtz2023aquila, henriet_quantum_2020}. This means the parameters will simply be the same for all qubits, \textit{i.e.} $\Omega_j(t)= \Omega(t)$, $\Delta_j(t)=\Delta(t)$ and $\phi_j(t)=\phi(t)$. Furthermore, we do not need to control the $\phi(t)$ term, since we are interested in performing annealing on the Ising model. Therefore, we choose $\phi(t)=0$. Then, the Hamiltonian takes the form,

\begin{align}
\frac{\mathcal{H}(t)}{\hbar} = \frac{\Omega(t)}{2} \sum_j X_j + \sum_{j =2}^n \sum_{k=1}^{j-1} \frac{V_{jk}}{4} Z_j Z_k - \frac14\sum_{j=1}^{n} \sum_{k\neq j} V_{jk} Z_j   +\frac{\Delta(t)}{2} \sum_{j} Z_j.
\end{align}

Now, we can use this equation to obtain the coefficients of Ising model,

\begin{equation}
    J_{ij} = \frac{V_{ij}}{4} = \frac{C_6}{4|\vec{r}_i-\vec{r}_j|^6},
\label{eq:ap-coupling-terms}
\end{equation}

\begin{align}
h_i = -\frac14\sum_{i\neq j}^{n} V_{ij} +\frac{\Delta(t)}{2} = -\sum_{i\neq j}^{n} J_{ij} +\frac{\Delta(t)}{2}.
\label{eq:ap-field-terms}
\end{align}

\subsection{\label{app:schedule} Comments on programming the device}

We observe that the field terms ($h_i$) appear to be dependent on the coupling terms ($J_{ij}$), as seen in Eq.~\ref{eq:ap-field-terms}. This is a result of not having local control of the field terms. To negate the influence of this, we choose the final $\Delta$ in the annealing schedule to be two times the average value of all couplings. Since it's not possible to cancel all terms, we try to do it on \textit{average}. Therefore, when setting the annealing schedule the final value of $\Delta$ will always be computed based on the chosen atom positions.

\subsection{\label{app:quera-atom} Sampling random Hamiltonians}

We need to pay attention to setting the scale correctly on the neutral-atom system. For our experiments we need Ising Hamiltonians with randomly sampled couplings. On a neutral atom system, this is achieved by choosing atom positions.

\begin{figure}[!t]
    \centering
    \includegraphics[width=\textwidth]{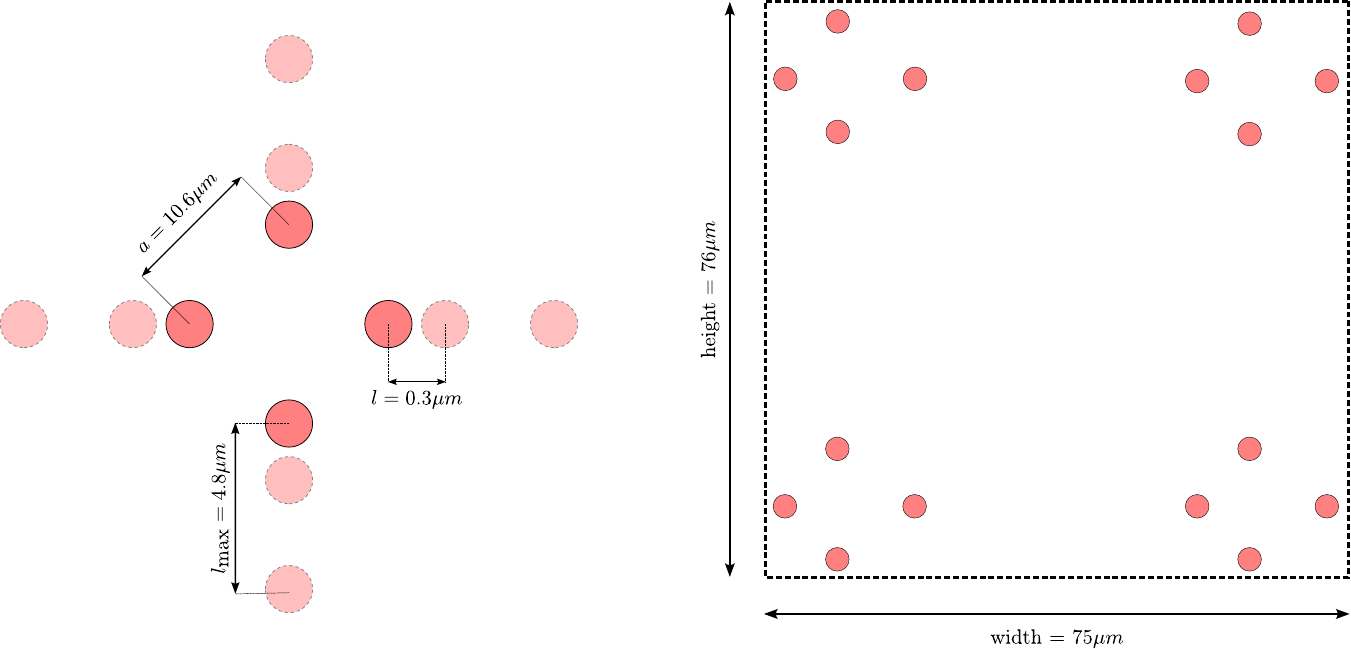}
    \caption{{\bf An illustration of the atom placement for the neutral atom experiments.} The atoms are placed on the edges of a $10.6 \mu m \times 10.6 \mu m$ square. The couplings formed by this arrangement forms the base Hamiltonian. Then, all atoms are positioned along their axis randomly up to $4.8 \mu m$ with $0.3 \mu m$ intervals. The set of random Hamiltonians forms part of the dataset. Next, these four atoms are tiled to the edges of the computing platform to maximize number of samples. Finally, each Hamiltonian is sampled on hardware and the samples are used the complete the dataset. }
    \label{fig:quera-setup}
\end{figure}

We would like to have $max(J_{ij}) \leq 1$ to follow the experiments on superconducting hardware. Putting this value in Eq.~\ref{eq:ap-coupling-terms} yields the minimum distance as $\approx 10.52$. Then, we define a base Hamiltonian. The base Hamiltonian has four atoms on the vertices of a square with the side length $10.6 \mu m$. Then, each atom is set to have one degree of freedom, along the diagonal of this square. Their positions are sampled randomly from a uniform distribution between $[0.0, 4.8]~\mu m$ with a step size is $0.3\mu m$. The step size is chosen to stay within the allowed precision of the hardware as described in Ref.~\cite{wurtz2023aquila}. We tile the four atom system to the corners of the computing plane in order to maximize the samples per execution. We limit the tiling only to the corners in order to avoid inter-subsystem interactions which may arise as subsystems come closer. This setup is illustrated in Fig.~\ref{fig:quera-setup}.

\begin{figure}[!t]
    \centering
    \includegraphics[width=0.5\linewidth]{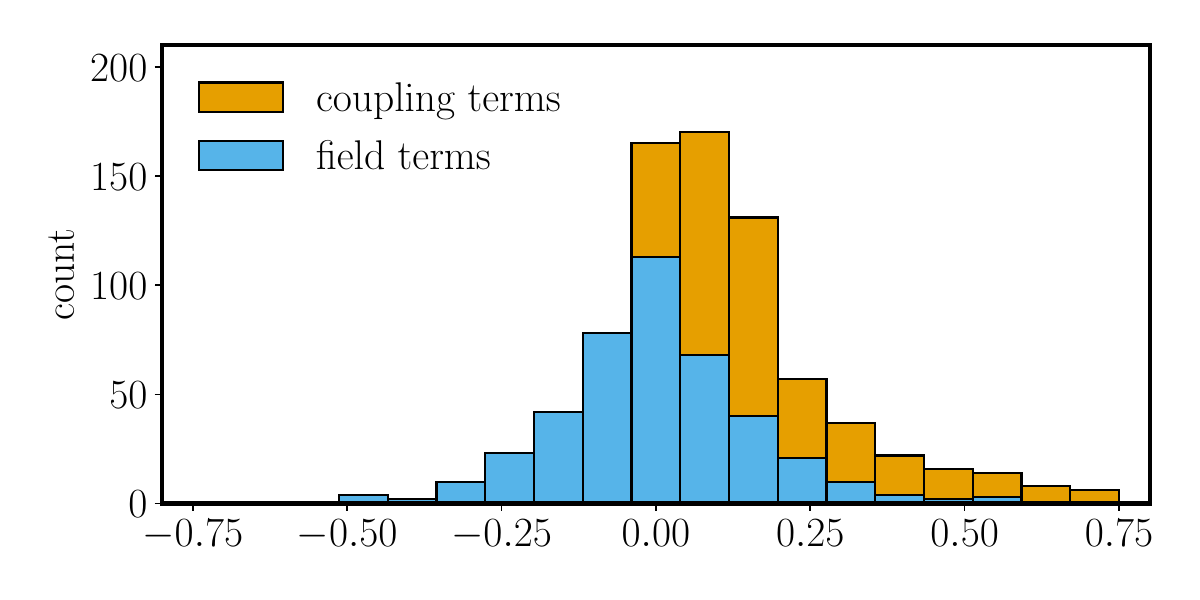}
    \caption{Histogram of fields and couplings of the randomly sampled 105 Hamiltonian used in the QuEra experiments. The field ($h_i$) and coupling ($J_{ij}$) terms are computed using \eqref{eq:ap-field-terms} and \eqref{eq:ap-coupling-terms} respectively.}
    \label{fig:enter-label}
\end{figure}

\begin{equation}
    \mu_{out}(\vec{\sigma}|\vec{\Theta}^{in}) = \frac{1}{Z} \exp\left(\sum_{i = 1}^n \sum_{j = i+1}^n J_{ij}^{out}(\vec{\Theta}^{in})\sigma_i \sigma_j + \sum_{i = 1}^n h_i^{out}(\vec{\Theta}^{in}) \sigma_i\right).
\end{equation}
\subsection {Technical details of DWave experiments}

\begin{table}[h!]
\centering
\begin{tabular}{lccccccc}
\toprule
 Experiment & Device & scale & offset & QPU anneal time & SRT & Qubit id  \\ 
\midrule
Fig 1 & 2000Q\_LANL & 1 & 0 & $5\mu$s &  1000 &Refer figure\\
Fig 2 & Advantage 4.1 & 1 & 0 & $5\mu$s &  False &Tiled\\
Fig 2 & Aquila & N/A & N/A & $3.5\mu$s &  N/A & N/A\\
Fig 3 & Advantage 4.1 & 1 & 0 & $5\mu$s &  False &[61,75,3120,3135]  \\
Fig 3 & 2000Q\_LANL & 1 & 0.0 & $5\mu$s &  1000  &  [304, 305, 308, 309]\\
\bottomrule
\end{tabular}
\caption{}
\label{table:example_transposed}
\end{table}

\section{Learned parameters}
\ca
\begin{figure}[!t]
    \centering
    \begin{minipage}[c]{.49\textwidth}
    \subfloat[\label{fig:1} Parameters of $F_{1}$ on 2000Q.]
    {\includegraphics[width=\textwidth]{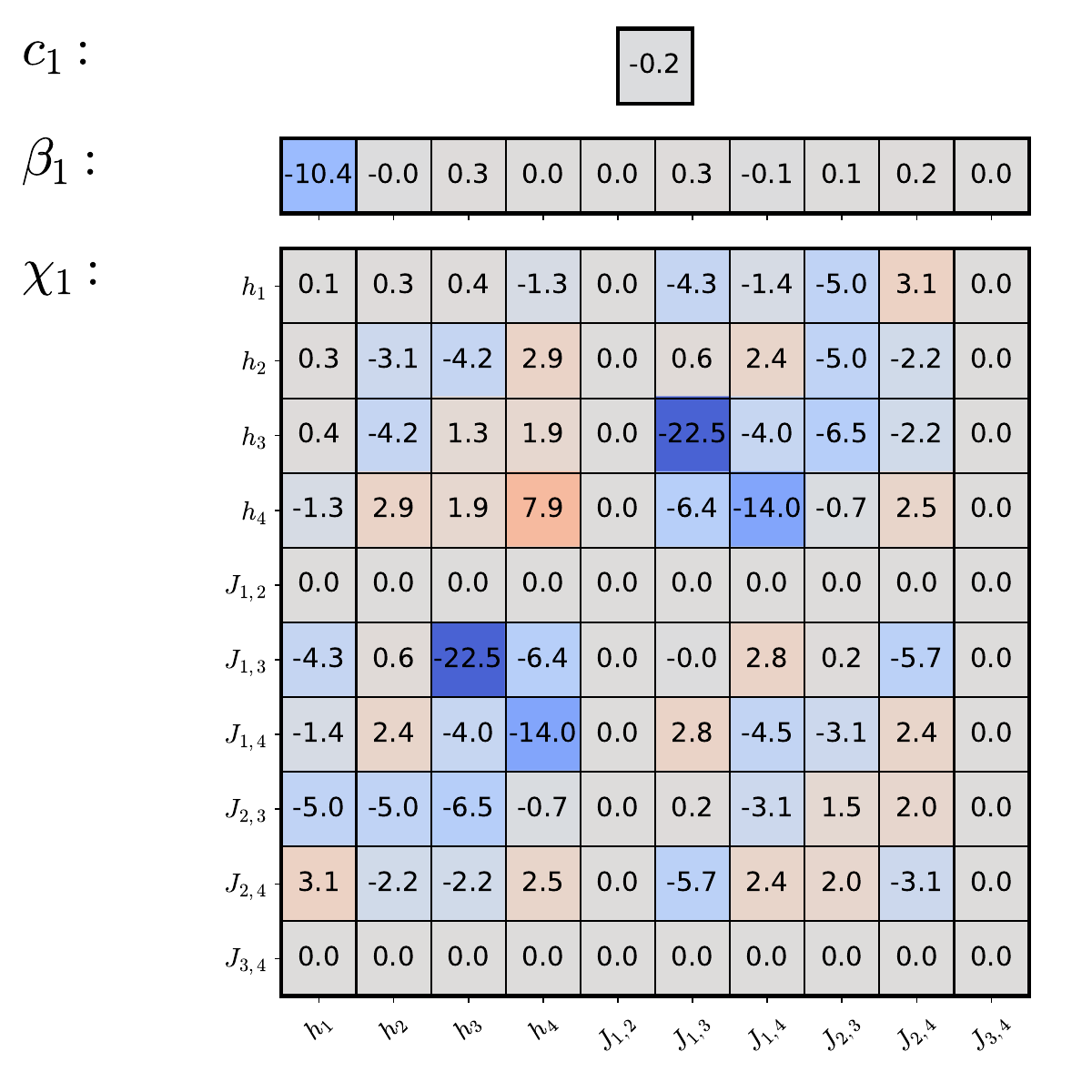}}
    \end{minipage}%
    \hfill%
    \begin{minipage}[c]{.49\textwidth}
    \subfloat[\label{fig:2} Parameters of $F_{3,4}$  on 2000Q.]
    {\includegraphics[width=\textwidth]{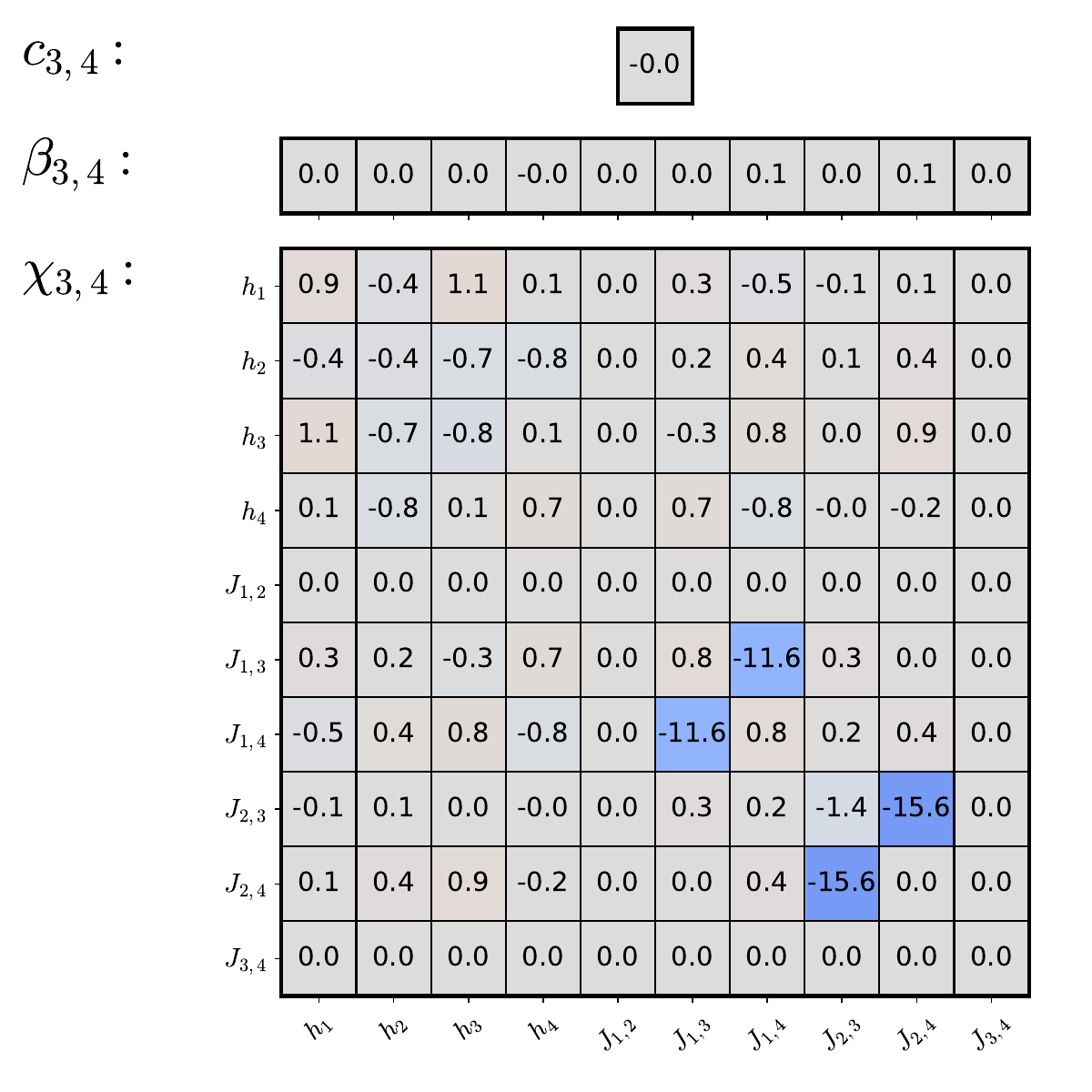}}
    \end{minipage}
    \begin{minipage}[c]{.49\textwidth}
    \subfloat[\label{fig:1} Parameters of $F_{1}$ on Advantage.]
    {\includegraphics[width=\textwidth]{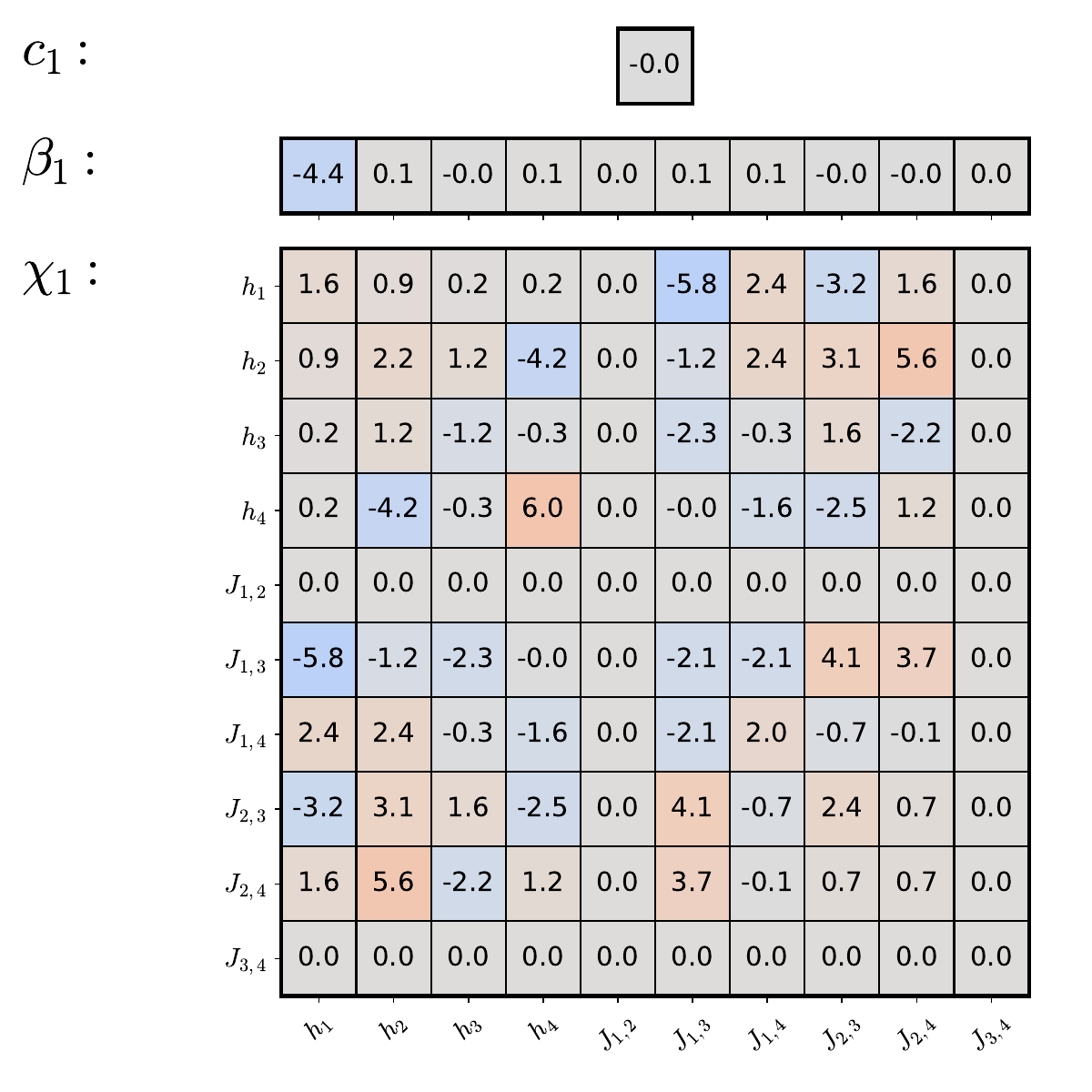}}
    \end{minipage}%
    \hfill%
    \begin{minipage}[c]{.49\textwidth}
    \subfloat[\label{fig:2} Parameters of $F_{3,4}$  on Advantage.]
    {\includegraphics[width=\textwidth]{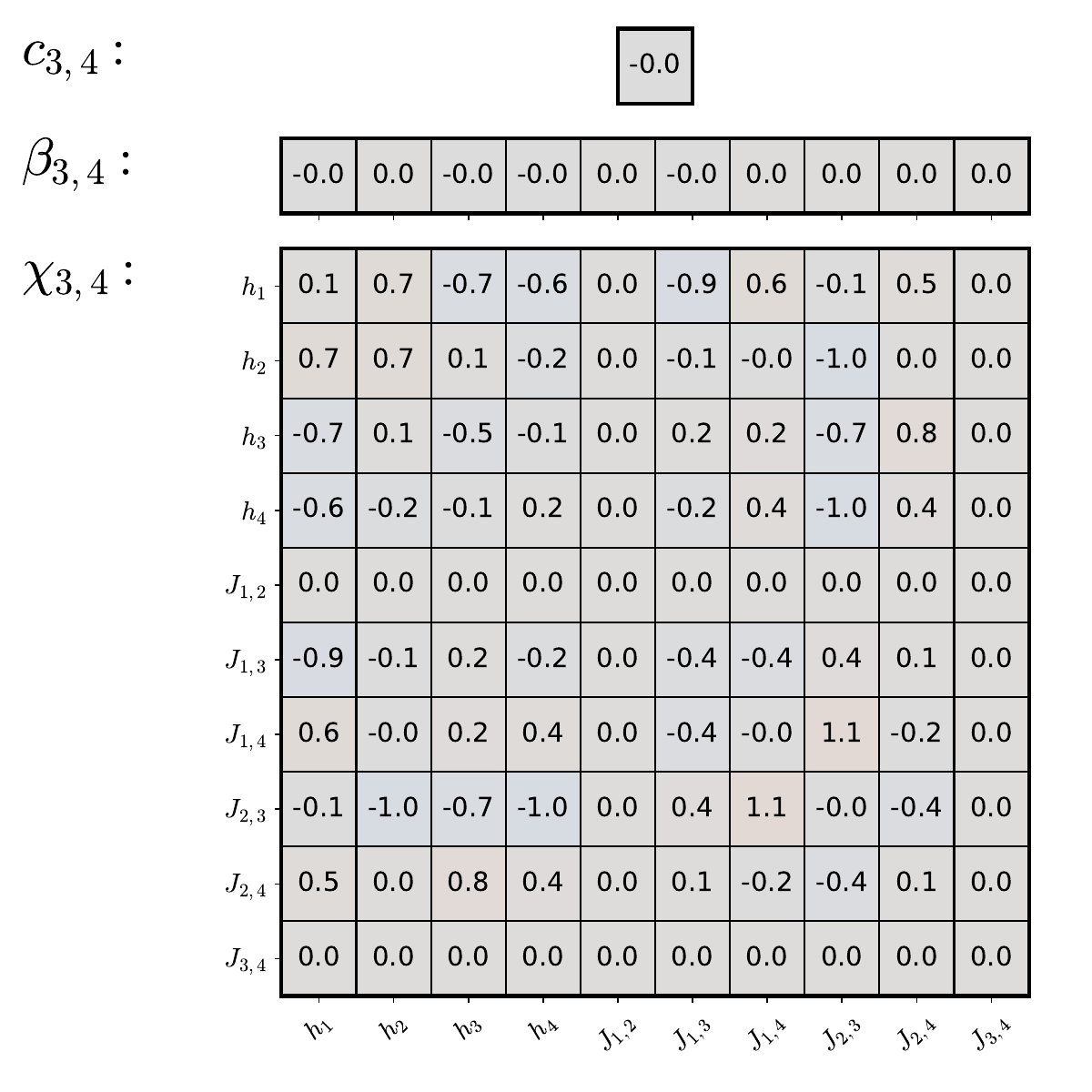}}
    \end{minipage}
    \caption{{\bf Parameters of the learned response function that corresponds to the \textit{non-tiled} 2000Q and Advantage experiments.} (a,c) Learned parameters of the $F_{1}$ response function. (b,d) Learned parameters of the $F_{3,4}$ response function. \ctcomm{I'm putting these figures here for now just as placeholder. We can plot them in any way you prefer. Please do comment on the style.} }
    \label{fig:learnerd-parameters}
\end{figure}
\cb

Recall the response function definitions,

\begin{align}
F_{ij}(\vec{\Theta}^{in}, \vec{\Phi}_{ij}) = c_{ij} +  ( \vec{\Theta}^{in} )^T \vec{\beta}_{ij} + ( \vec{\Theta}^{in} )^T {\chi}_{ij} \vec{\Theta}^{in},\\ 
F_{i} (\vec{\Theta}^{in}, \vec{\Phi}_{i}) = c_{i} +  ( \vec{\Theta}^{in} )^T \vec{\beta}_{i} + ( \vec{\Theta}^{in} )^T {\chi}_{i} \vec{\Theta}^{in}.
\end{align}

The values of these specific coefficients learned from D-Wave and QuEra machines are given below for completeness. \figurename 6 and 7 show the responses learned in Experiment 3. While \figurename 8 and 9 correspond to the responses learned in Experiment 2.

\begin{figure}[htp]
  \centering
  \begin{minipage}{0.45\linewidth}
    \centering
    \includegraphics[width=\linewidth]{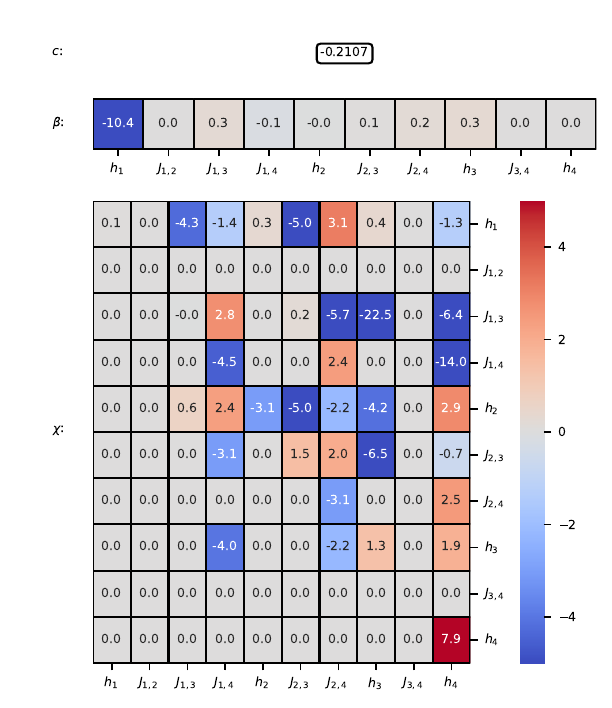}
    \subcaption{$F_1$}
  \end{minipage}\hfill
  \begin{minipage}{0.45\linewidth}
    \centering
    \includegraphics[width=\linewidth]{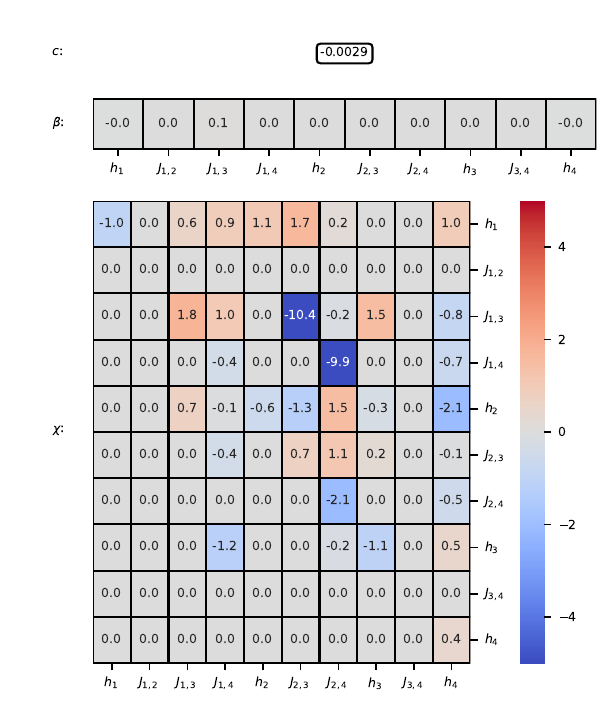}
    \subcaption{$F_{12}$}
  \end{minipage}
  
  \vspace{1em} 
  
  \begin{minipage}{0.45\linewidth}
    \centering
    \includegraphics[width=\linewidth]{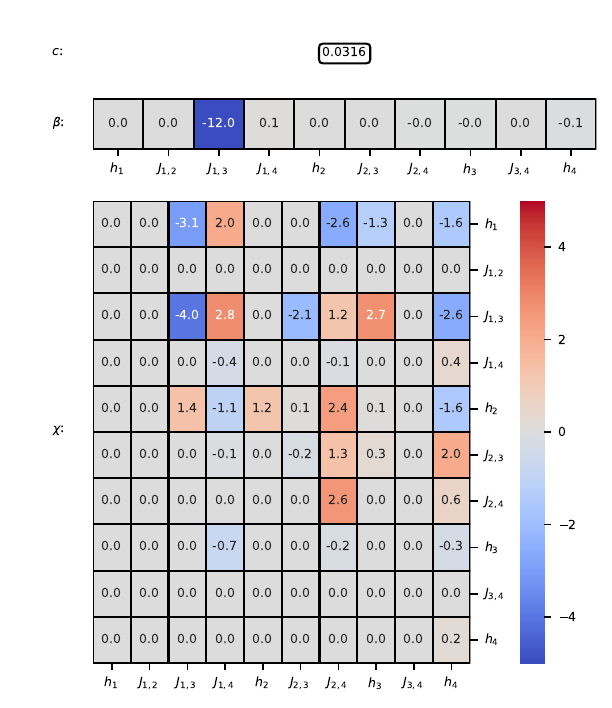}
    \subcaption{$F_{13}$}
  \end{minipage}\hfill
  \begin{minipage}{0.45\linewidth}
    \centering
    \includegraphics[width=\linewidth]{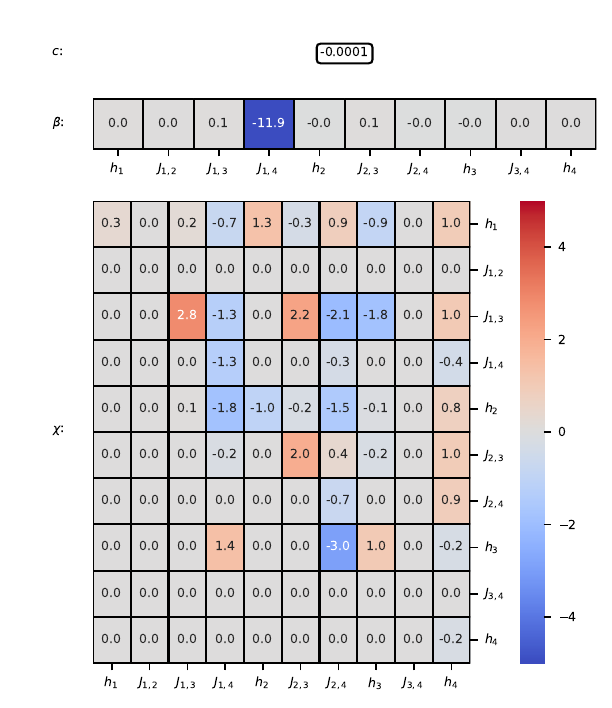}
    \subcaption{$F_{14}$}
  \end{minipage}
  
  \vspace{1em} 
  
  \caption{Learned response function parameters for qubits 304,305,308,309 in DWave 2000Q machine  (Part 1)}
 \end{figure}

\begin{figure}[htp]
  \ContinuedFloat 
  \centering
  
  \begin{minipage}{0.45\linewidth}
    \centering
    \includegraphics[width=\linewidth]{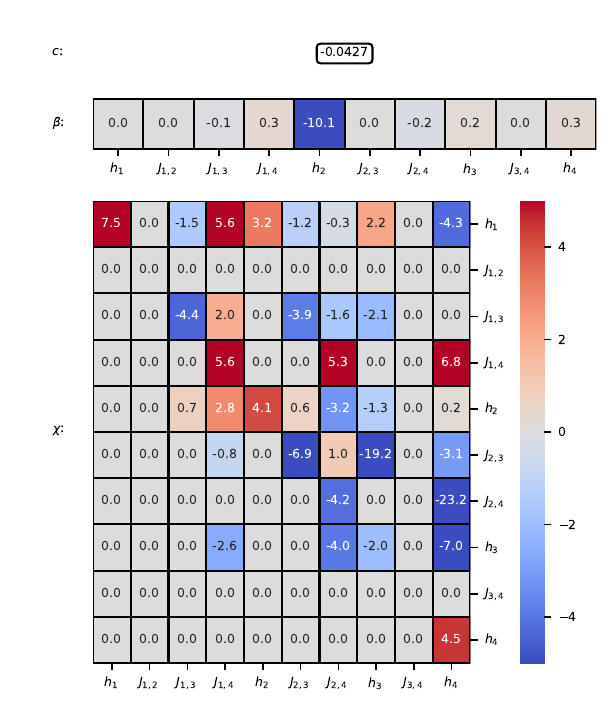}
    \subcaption{$F_2$}
  \end{minipage}\hfill
  \begin{minipage}{0.45\linewidth}
    \centering
    \includegraphics[width=\linewidth]{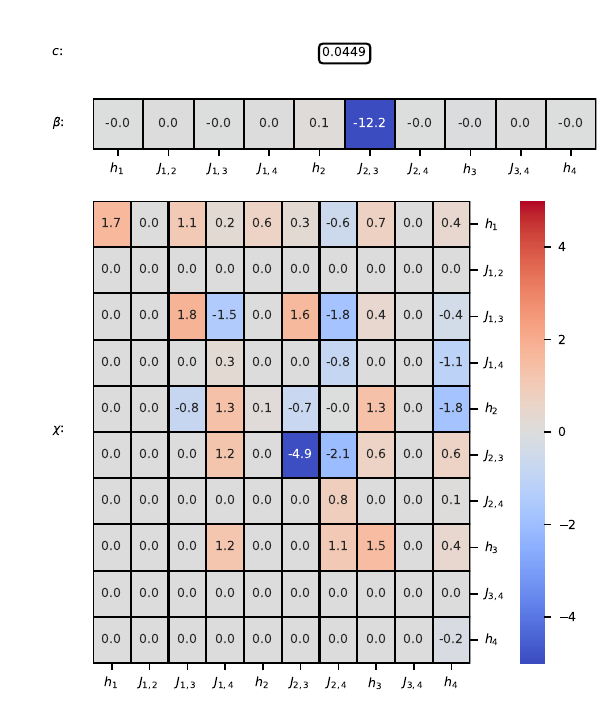}
    \subcaption{$F_{23}$}
  \end{minipage}
  
  \begin{minipage}{0.45\linewidth}
    \centering
    \includegraphics[width=\linewidth]{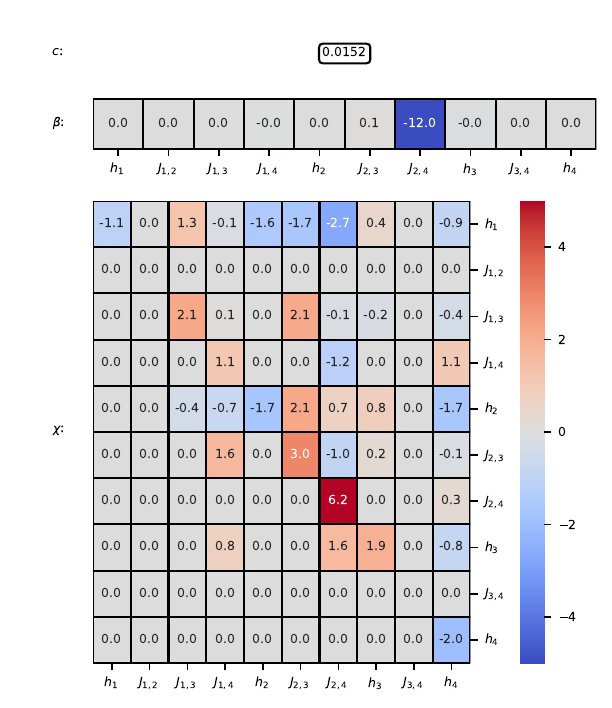}
    \subcaption{$F_{24}$}
  \end{minipage}\hfill
  \begin{minipage}{0.45\linewidth}
    \centering
    \includegraphics[width=\linewidth]{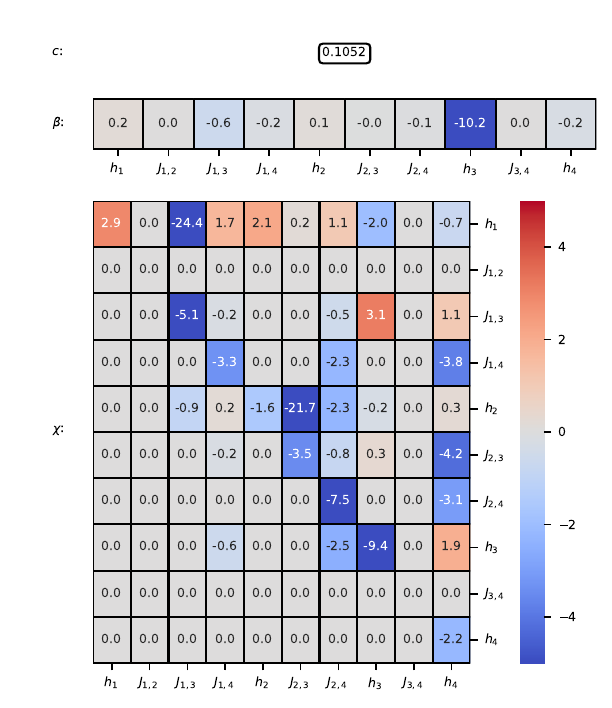}
    \subcaption{$F_{3}$}
  \end{minipage}
  
  \vspace{1em} 
    
  \caption[]{Learned response function parameters for qubits 304,305,308,309 in DWave 2000Q machine (Part 2, continued) }
\end{figure}

\begin{figure}[htp]
  \ContinuedFloat 
  \centering

  \begin{minipage}{0.45\linewidth}
    \centering
    \includegraphics[width=\linewidth]{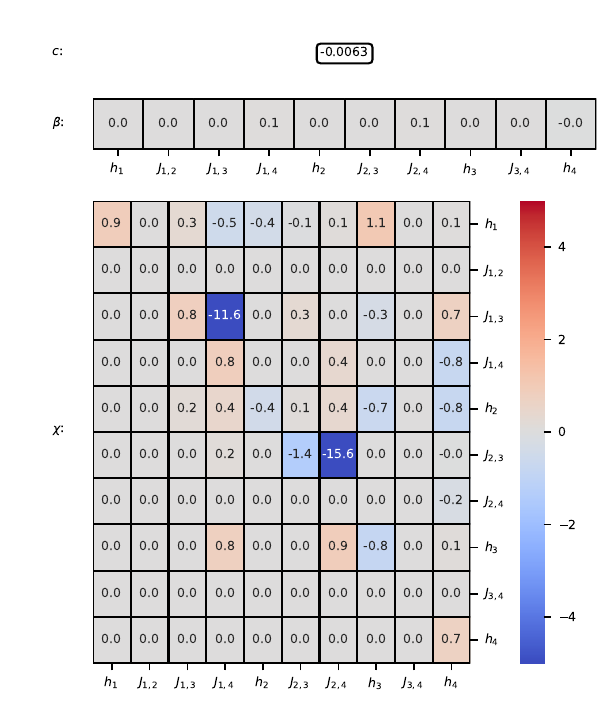}
    \subcaption{$F_{34}$}
  \end{minipage}\hfill
  \begin{minipage}{0.45\linewidth}
    \centering
    \includegraphics[width=\linewidth]{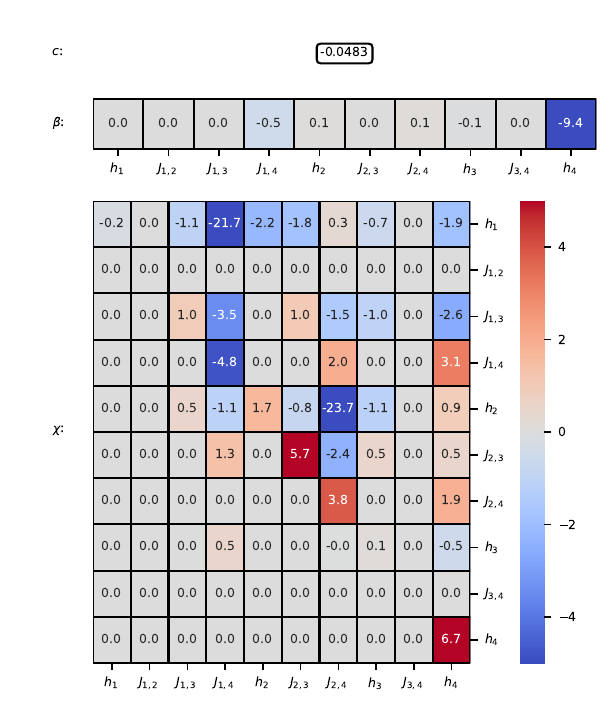}
    \subcaption{$F_{4}$}
  \end{minipage}
  \caption[]{Learned response function parameters for qubits 304,305,308,309 in DWave 2000Q machine (Part 3, continued) }
\end{figure}

\begin{figure}[htp]
  \centering
  \begin{minipage}{0.45\linewidth}
    \centering
    \includegraphics[width=\linewidth]{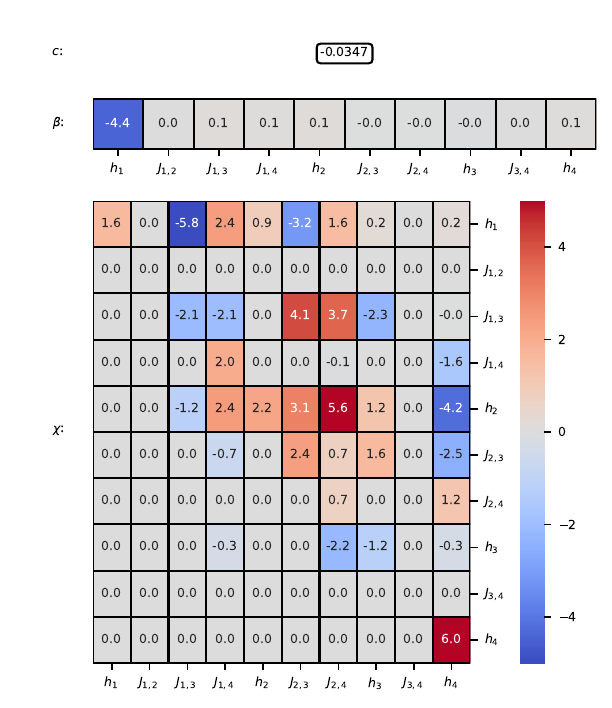}
    \subcaption{$F_1$}
  \end{minipage}\hfill
  \begin{minipage}{0.45\linewidth}
    \centering
    \includegraphics[width=\linewidth]{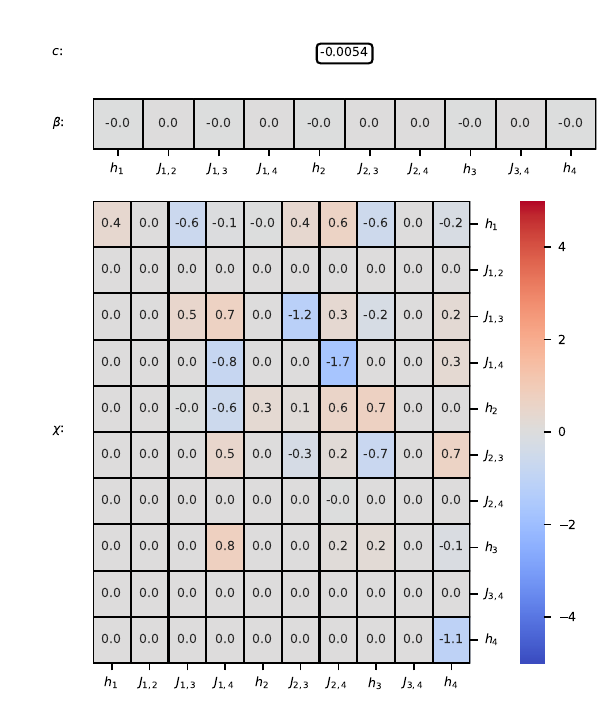}
    \subcaption{$F_{12}$}
  \end{minipage}

  \vspace{1em} 
  
  \begin{minipage}{0.45\linewidth}
    \centering
    \includegraphics[width=\linewidth]{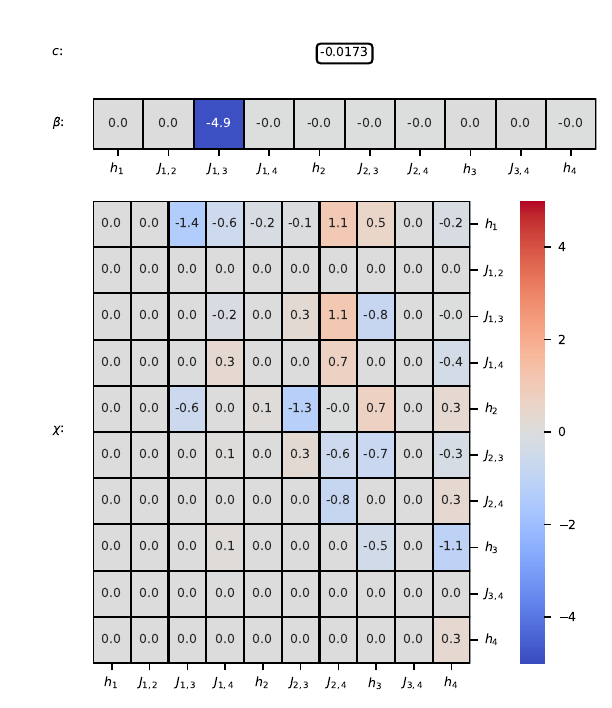}
    \subcaption{$F_{13}$}
  \end{minipage}\hfill
  \begin{minipage}{0.45\linewidth}
    \centering
    \includegraphics[width=\linewidth]{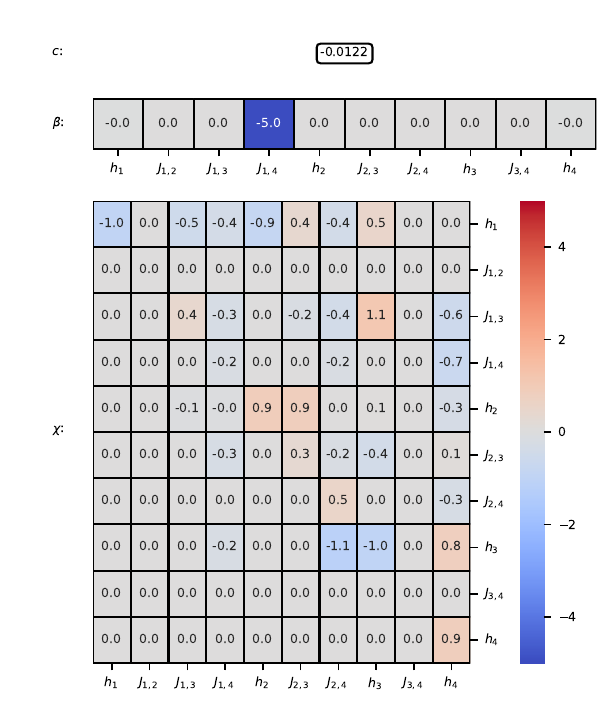}
    \subcaption{$F_{14}$}
  \end{minipage}
  
  \vspace{1em} 
  \caption{ Learned response function parameters for qubits 61,75,3120,3135 in DWave Advantage machine (Part 1)}
 \end{figure}

\begin{figure}[htp]
  \ContinuedFloat 
  \centering
  
  \begin{minipage}{0.45\linewidth}
    \centering
    \includegraphics[width=\linewidth]{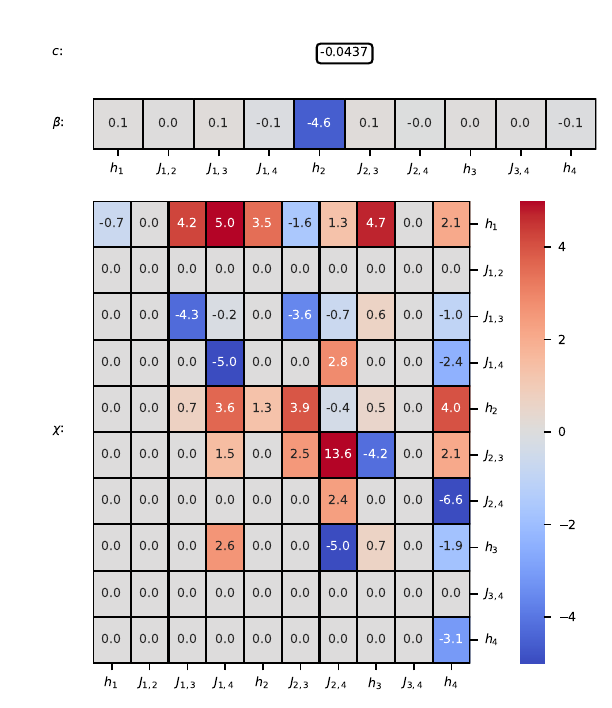}
    \subcaption{$F_{2}$}
  \end{minipage}\hfill
  \begin{minipage}{0.45\linewidth}
    \centering
    \includegraphics[width=\linewidth]{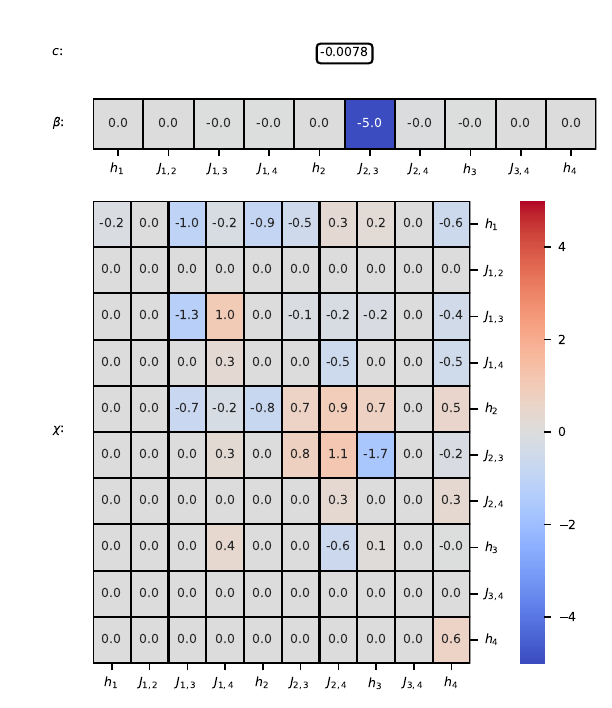}
    \subcaption{$F_{23}$}
  \end{minipage}
  
  \begin{minipage}{0.45\linewidth}
    \centering
    \includegraphics[width=\linewidth]{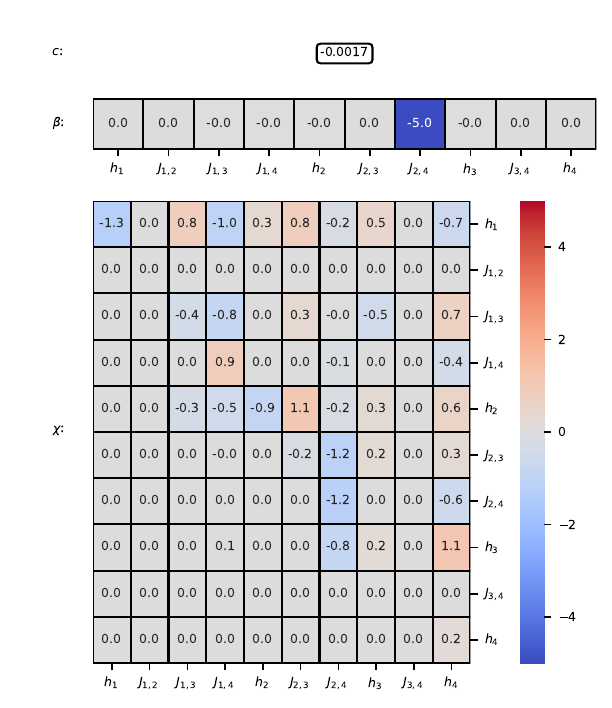}
    \subcaption{$F_{24}$}
  \end{minipage}\hfill
  \begin{minipage}{0.45\linewidth}
    \centering
    \includegraphics[width=\linewidth]{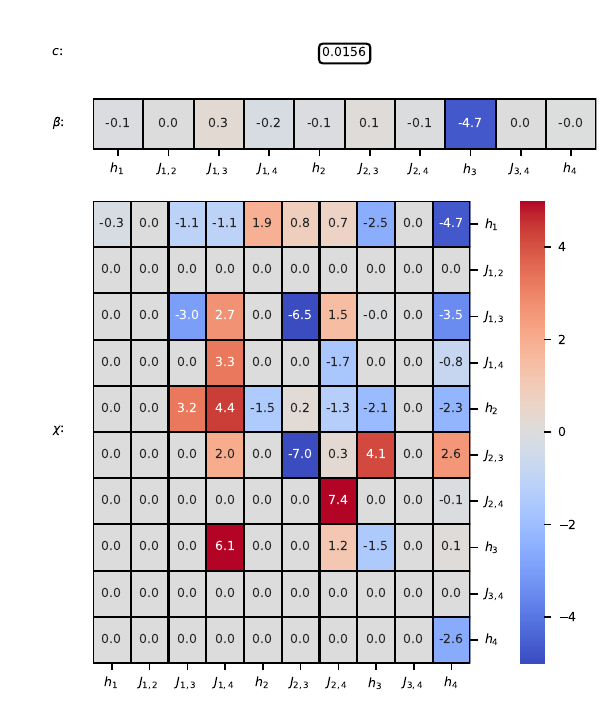}
    \subcaption{$F_{33}$}
  \end{minipage}
  
  \vspace{1em} 
    
  \caption[]{Learned response function parameters for qubits 61,75,3120,3135 in DWave Advantage machine (Part 2, continued)}
\end{figure}

\begin{figure}[htp]
  \ContinuedFloat 
  \centering

  \begin{minipage}{0.45\linewidth}
    \centering
    \includegraphics[width=\linewidth]{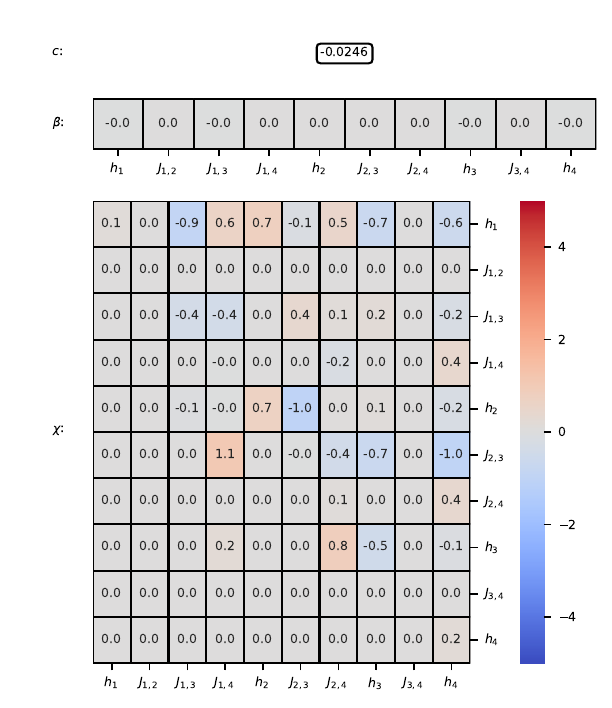}
    \subcaption{$F_{34}$}
  \end{minipage}\hfill
  \begin{minipage}{0.45\linewidth}
    \centering
    \includegraphics[width=\linewidth]{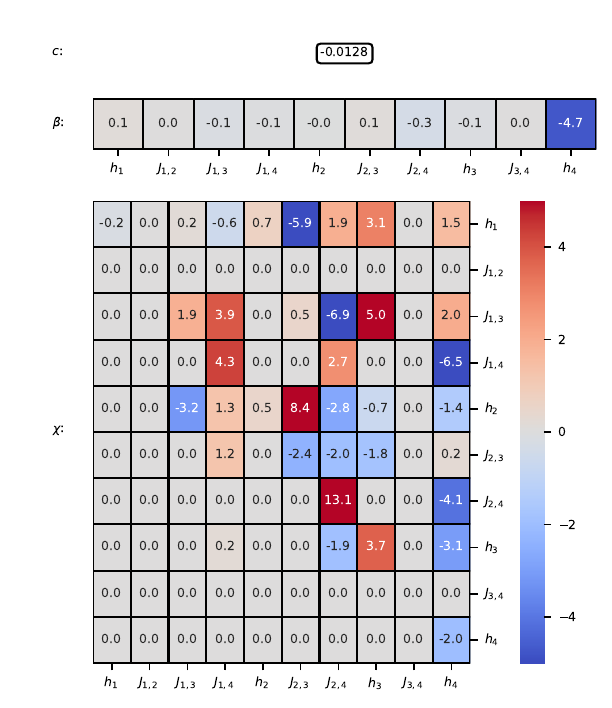}
    \subcaption{$F_{4}$}
  \end{minipage}
  \caption[]{ Learned response function parameters for qubits 61,75,3120,3135 in DWave Advantage machine (Part 3, continued)}
\end{figure}

\begin{figure}[htp]
  \centering
  \begin{minipage}{0.45\linewidth}
    \centering
    \includegraphics[width=\linewidth]{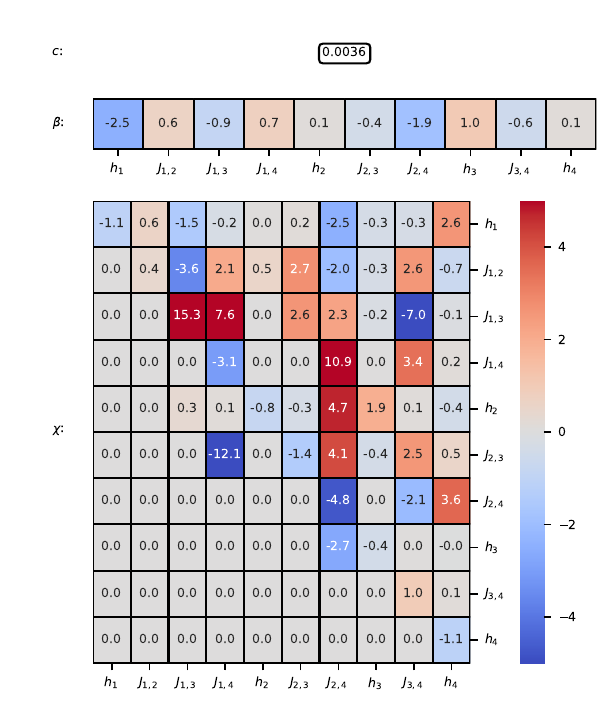}
    \subcaption{$F_1$}
  \end{minipage}\hfill
  \begin{minipage}{0.45\linewidth}
    \centering
    \includegraphics[width=\linewidth]{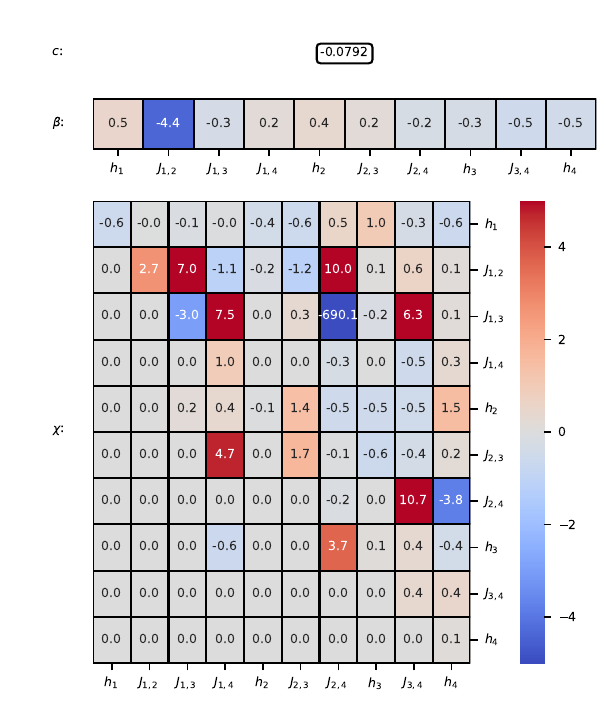}
    \subcaption{$F_{12}$}
  \end{minipage}
  
  \vspace{1em} 
  
  \begin{minipage}{0.45\linewidth}
    \centering
    \includegraphics[width=\linewidth]{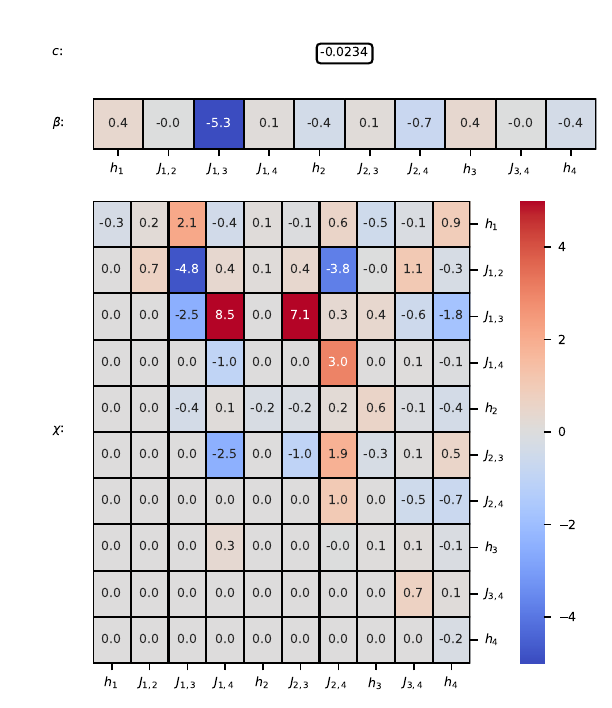}
    \subcaption{$F_{13}$}
  \end{minipage}\hfill
  \begin{minipage}{0.45\linewidth}
    \centering
    \includegraphics[width=\linewidth]{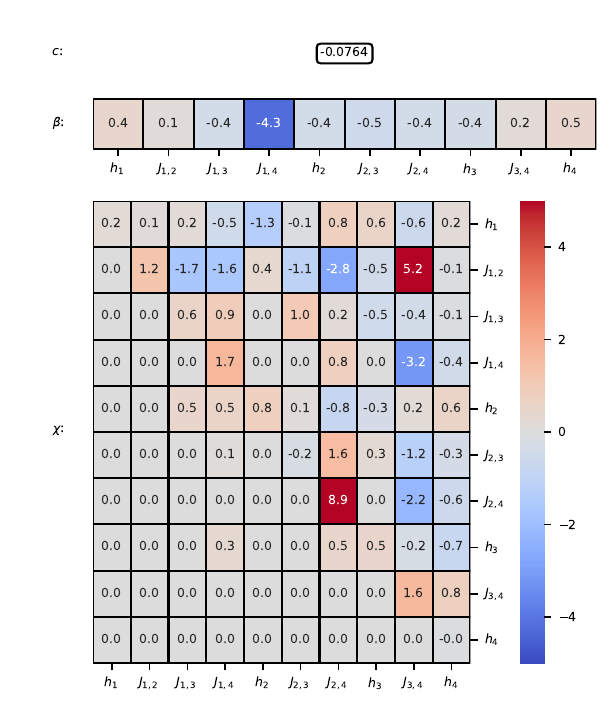}
    \subcaption{$F_{14}$}
  \end{minipage}
  
  \vspace{1em} 
  \caption{Average response DWave Advantage 4.1 (Part 1)}
 \end{figure}

\begin{figure}[htp]
  \ContinuedFloat 
  \centering
  
  \begin{minipage}{0.45\linewidth}
    \centering
    \includegraphics[width=\linewidth]{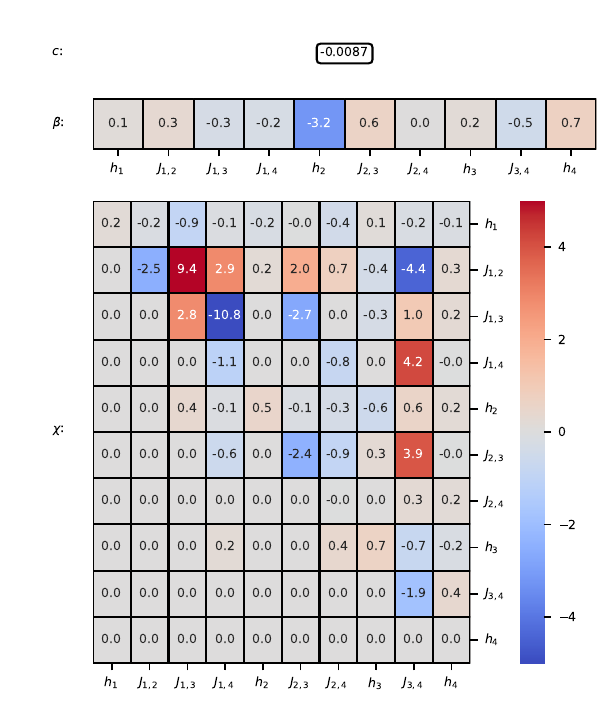}
    \subcaption{$F_{2}$}
  \end{minipage}\hfill
  \begin{minipage}{0.45\linewidth}
    \centering
    \includegraphics[width=\linewidth]{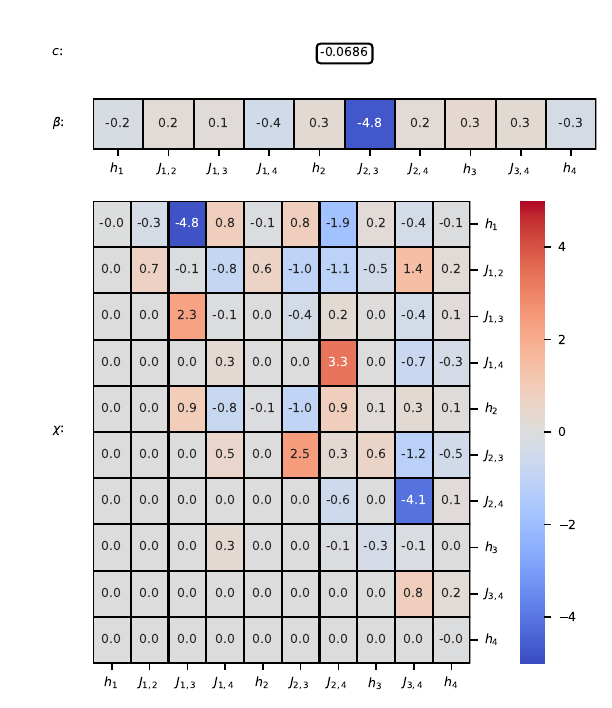}
    \subcaption{$F_{23}$}
  \end{minipage}
  
  \begin{minipage}{0.45\linewidth}
    \centering
    \includegraphics[width=\linewidth]{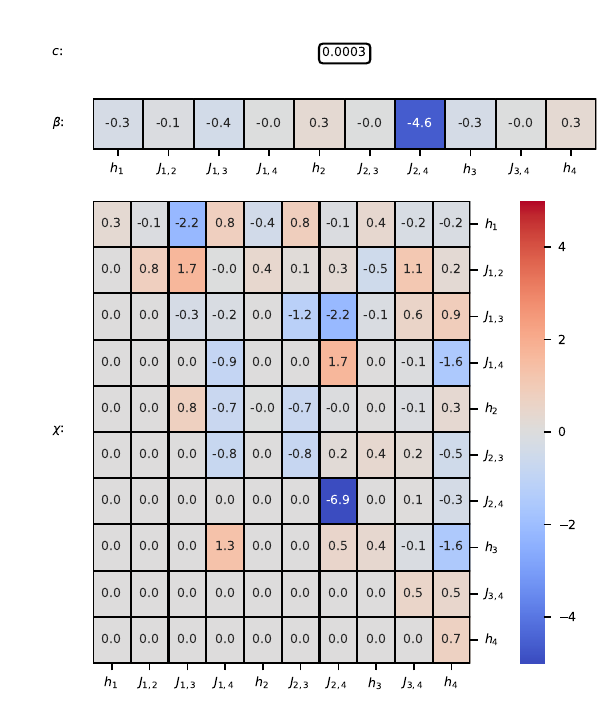}
    \subcaption{$F_{24}$}
  \end{minipage}\hfill
  \begin{minipage}{0.45\linewidth}
    \centering
    \includegraphics[width=\linewidth]{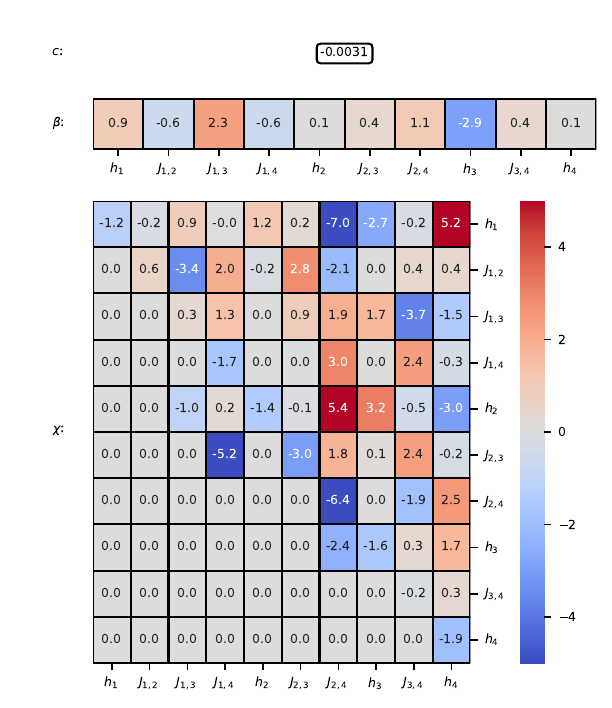}
    \subcaption{$F_3$}
  \end{minipage}
  
  \vspace{1em} 
    
  \caption[]{ Average response DWave Advantage 4.1 (Part 2, continued)}
\end{figure}

\begin{figure}[htp]
  \ContinuedFloat 
  \centering

  \begin{minipage}{0.45\linewidth}
    \centering
    \includegraphics[width=\linewidth]{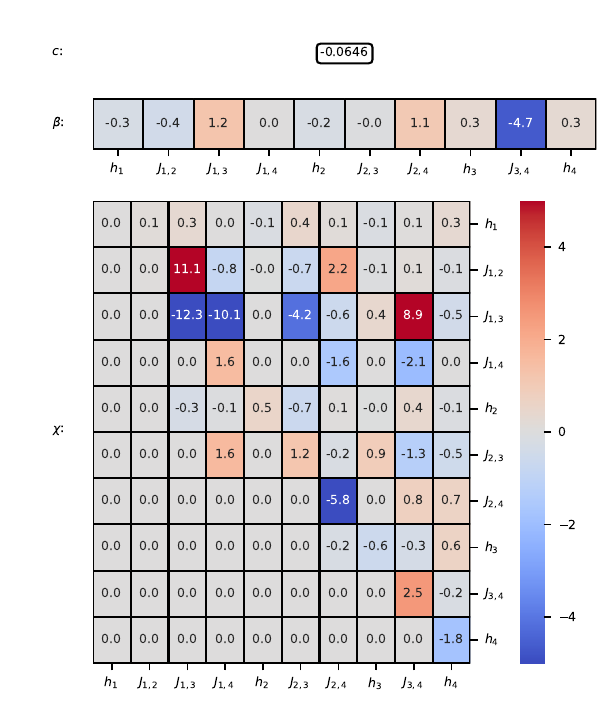}
    \subcaption{$F_{34}$}
  \end{minipage}\hfill
  \begin{minipage}{0.45\linewidth}
    \centering
    \includegraphics[width=\linewidth]{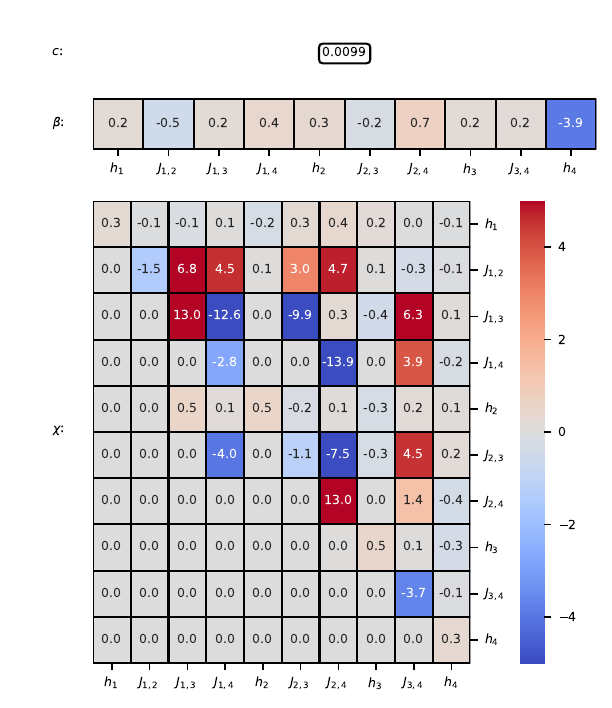}
    \subcaption{$F_4$}
  \end{minipage}
  \caption[]{ Average response DWave Advantage 4.1(Part 3, continued)}
\end{figure}

\begin{figure}[htp]
  \centering
  \begin{minipage}{0.45\linewidth}
    \centering
    \includegraphics[width=\linewidth]{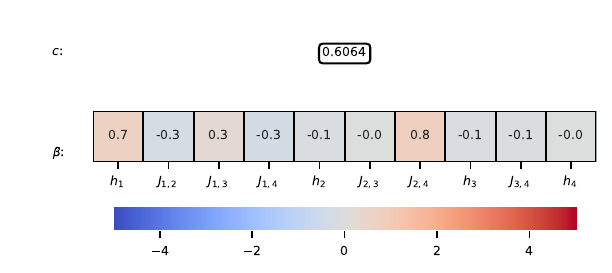}
    \subcaption{$F_1$}
  \end{minipage}\hfill
  \begin{minipage}{0.45\linewidth}
    \centering
    \includegraphics[width=\linewidth]{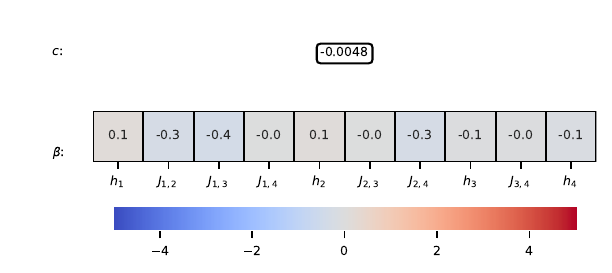}
    \subcaption{$F_{12}$}
  \end{minipage}
  
  \vspace{1em} 
  
  \begin{minipage}{0.45\linewidth}
    \centering
    \includegraphics[width=\linewidth]{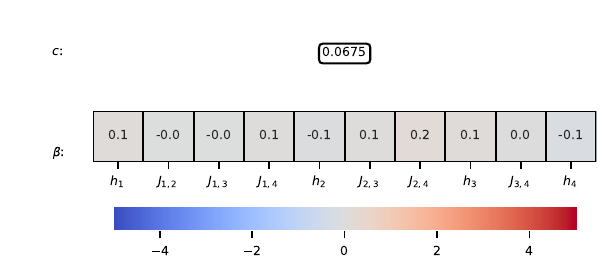}
    \subcaption{$F_{13}$}
  \end{minipage}\hfill
  \begin{minipage}{0.45\linewidth}
    \centering
    \includegraphics[width=\linewidth]{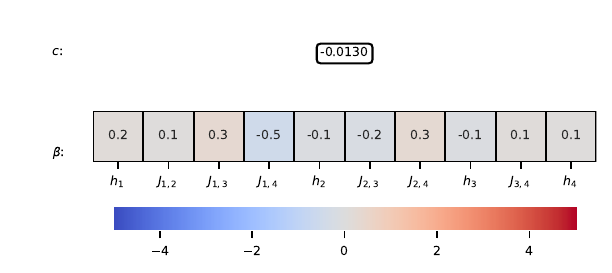}
    \subcaption{$F_{14}$}
  \end{minipage}
  
  \caption{Average response QuEra Aquila (Part 1}
  \vspace{1em} 
 \end{figure}

\begin{figure}[htp]
  \ContinuedFloat 
  \centering
  
  \begin{minipage}{0.45\linewidth}
    \centering
    \includegraphics[width=\linewidth]{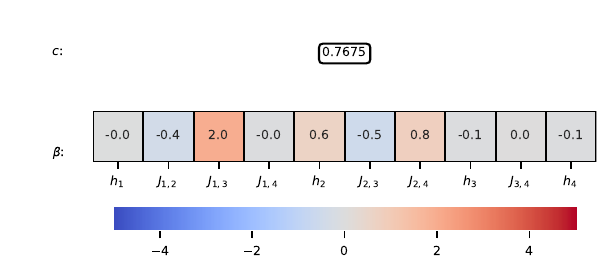}
    \subcaption{$F_{2}$}
  \end{minipage}\hfill
  \begin{minipage}{0.45\linewidth}
    \centering
    \includegraphics[width=\linewidth]{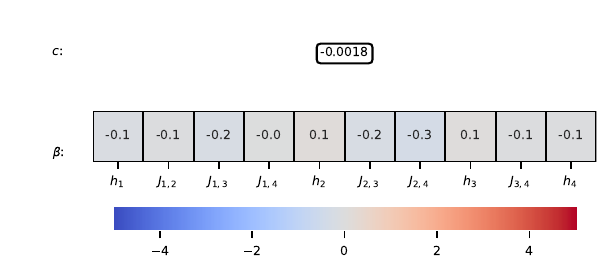}
    \subcaption{$F_{23}$}
  \end{minipage}
  
  \begin{minipage}{0.45\linewidth}
    \centering
    \includegraphics[width=\linewidth]{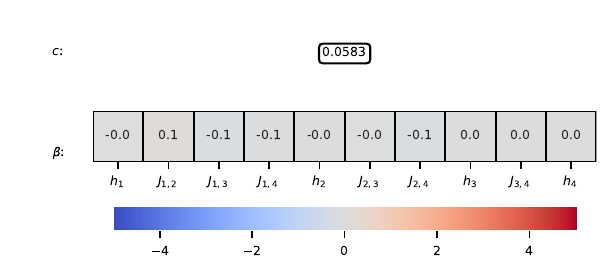}
    \subcaption{$F_{24}$}
  \end{minipage}\hfill
  \begin{minipage}{0.45\linewidth}
    \centering
    \includegraphics[width=\linewidth]{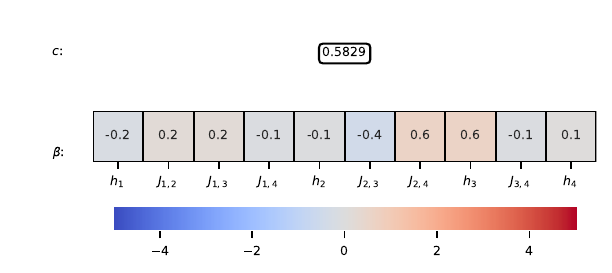}
    \subcaption{$F_{3}$}
  \end{minipage}
  \begin{minipage}{0.45\linewidth}
    \centering
    \includegraphics[width=\linewidth]{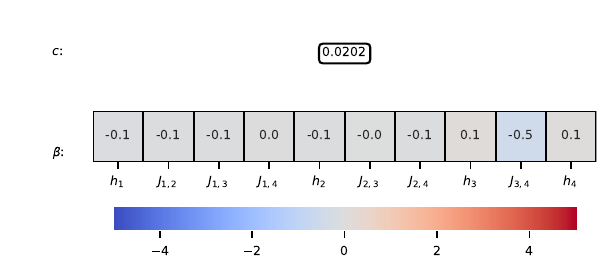}
    \subcaption{$F_{34}$}
  \end{minipage}\hfill
  \begin{minipage}{0.45\linewidth}
    \centering
    \includegraphics[width=\linewidth]{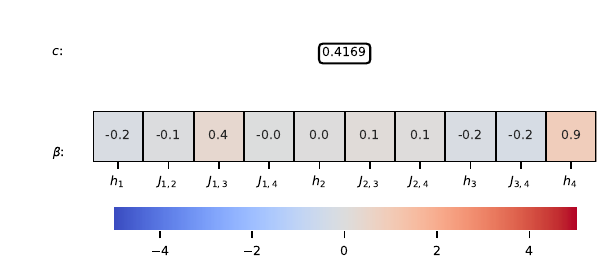}
    \subcaption{$F_4$}
  \end{minipage}
  \caption{Average response QuEra Aquila (Part 2)}
\end{figure}

\end{document}